\documentclass[%
amsmath,
amssymb,
prb,
twocolumn,
groupaddress,
]{revtex4-2}

\newcommand{\cgreen}{\color{teal}}  

\usepackage{graphicx}
\usepackage{dcolumn}
\usepackage{bm}
\usepackage[normalem]{ulem}
\usepackage[makeroom]{cancel}

\usepackage{verbatim}
\usepackage{svg}
\usepackage{caption}
\usepackage{ragged2e}
\usepackage{subcaption}

\usepackage{filecontents}

\newif\ifdraft
\draftfalse
\ifdraft {} \else \renewcommand{\sout}[1]{\unskip} \fi
\ifdraft {} \else \renewcommand{\cgreen}{\color{black}} \fi

\begin{document}

\title{
  Static and fluctuating zigzag order, 
  and possible signatures of Kitaev physics, 
  in torque measurements of $\alpha$-RuCl$_3$}

\author{S. Froude-Powers}
\author{Subin Kim}
\author{Jacob Gordon}
\author{Hae-Young Kee}
\author{Young-June Kim}
\author{S.R. Julian}\email{stephen.julian@utoronto.ca}
\affiliation{%
Department of Physics, University of Toronto,
Toronto, Ontario, M5S 1A7 Canada }

\date{December 23, 2023}

\begin{abstract}
We have measured magnetic torque on a $T_N=7$ K
  single crystal of $\alpha$-RuCl$_3$, 
  as a function of the 
  field angle in the $ab$-plane, 
  focusing on temperatures between 2 and 20 K 
  and fields from 0 to 9 T. 
We find a 
\sout{ rich spectrum of signals} 
  {\cgreen number of features}, many of which can be 
  classified by their angular periodicity. 
The sample shows
  an oscillation with a period of 180$^\circ$ 
  (i.e.\ two-fold periodicity) 
  \sout{which we argue is due to residual strain
  within the crystal, rather than being intrinsic.}
\sout{In addition,}
{\cgreen 
 and} 
  within the magnetically ordered zigzag phase 
  there is a 60$^\circ$ period (i.e.\ six-fold) 
  sawtooth pattern, which can be explained by
  reorientation of the zigzag domains as the 
  crystal rotates in the applied field. 
{\cgreen
We argue that the six-fold sawtooth and the 
  two-fold sinusoidal signals 
  arise from distinct regions of the crystal.} 
Suppressing the zigzag order 
  with an applied field above $\sim8$ T at 
  low temperature, a 
  six-fold {\sl sinusoidal} signal remains, 
  suggesting that there is fluctuating zigzag order 
  in the putative field-induced quantum spin liquid state.
\sout{Finally, our key finding
  is a sharp, step-like feature that appears 
   at low temperature for fields just 
   above the zigzag phase 
   boundary, at the so-called B2-axes.
This is similar to theoretically predicted behaviour 
  for a state with Ising topological order, which is expected for 
  a Kitaev spin liquid in an applied magnetic field.}
{\cgreen
Finally, in testing theoretical results which predict a 
  torque response with divergent slope
  across C$_2$-preserving 
  $b$-axes (B1-axis), we find no features like that predicted for
  Ising topological order.
Instead we find features at low temperatures and fields
  just above the zigzag phase across the non-C$_2$-preserving
  $b$-axes (B2-axes). 
Interpretation of this feature is complicated 
  by the development of other similar 
  signatures nearby at slightly lower fields, 
  and by clear enhancement with thermal cycling.
Additionally, we contrast 
  the torque response of $T_N 
  \sim$ 7 K and 14 K samples.
}

\end{abstract}

\maketitle

\section{Introduction}
\label{sec-intro}
Quantum spin liquids have generated interest both as a
  new phase of condensed matter and for their potential 
  applications in the fields of quantum computing and 
  spintronics, where they offer a paradigm-shifting alternative 
  to traditional electronics. 
The recent incorporation of spin-orbit coupling 
  has played a key role 
  \cite{Rau2016,Rousochatzakis2023,Witczak2014,Takayama2021},
  by making {\sl topological} 
  quantum spin liquids theoretically possible. 
The analytically solvable Kitaev model,
  with bond-dependent Ising-type interactions of 
  $J_{\rm eff}=1/2$
  nearest-neighbours on a 2D honeycomb lattice, leads 
  to frustrated spins and an emergent topological 
  quantum spin liquid (QSL) \cite{Kitaev2006}.
With no applied magnetic field this QSL is found to
  host emergent excitations of gapless Majorana fermions 
  and gapped Z$_2$ vortices
  - novel excitations should they be found in  nature. 
The possibility of 
  realizing the Kitaev model in real materials 
  has aroused considerable interest (for reviews see  
  \cite{Hermanns2018,Takagi2019,Motome2020,Winter2016,Winter2017}),
  and of the various candidate materials 
  $\alpha$-RuCl$_3$, which is the 
  focus of this paper, has been the subject of 
  several experimental and theoretical studies.

\begin{figure}
    \centering
    \hspace*{-0.7cm}%
    \includegraphics{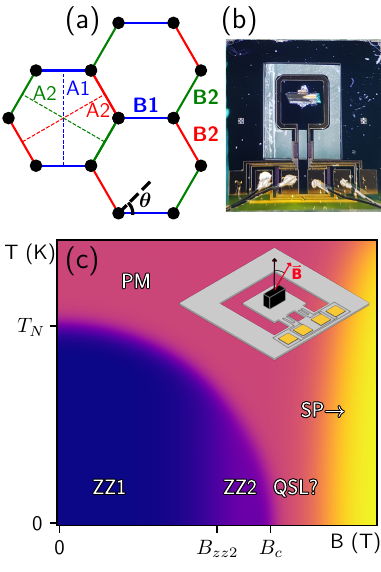}
    \caption{\justifying%
    (a)
      The three X,Y,Z Ru-Ru bond directions 
        of $\alpha$-RuCl$_3$: the $b$-axes are  
        B1 (the $C_2$ axis for a crystal with 
        monoclinic symmetry), and two equivalent 
        B2 axes.  
      For R$\overline{3}$ symmetry, B1 and 
        B2 are equivalent.
      The {\cgreen complementary} 
      \sout{complimentary} $a$-axes, 
        A1 and A2, are shown as dashed lines.
      The nomenclature is borrowed from Lampen-Kelley 
        et al. \cite{LampenKelley2018}. 
    (b)
      Top view of \sout{sample L7K} 
        {\cgreen our large $T_N \sim 7$ K sample (sample L7K)} 
        mounted on the Quantum Design torque chip P109A.
    (c) 
      Cartoon phase diagram of $\alpha$-RuCl$_3$.  
      Above the zigzag ordering temperature $T_N \simeq 7$ K, 
        $\alpha$-RuCl$_3$ is in the 
        \sout{classical paramagnetic phase (CPM)}
        {\cgreen
        paramagnetic (PM) phase.
        }
      At zero field, below $T_N$ $\alpha$-RuCl$_3$ 
        has zigzag 
        antiferromagnetic order (ZZ1). 
      Increasing the field beyond a threshold $B_{zz2}$ induces a new 
        zigzag ordering (ZZ2) \cite{Balz2019}. 
      Zigzag order is suppressed for 
        fields beyond $B_c \sim 7.5$ T, at high field reaching a 
        field-induced 
        \sout{quantum paramagnetic state (QPM).}
        {\cgreen
        spin-polarized (SP) state.
        }
      \sout{
      In the intermediate field phase that 
      lies between the ZZ2  
        and the }
        \sout{QPM} 
        \sout{\cgreen
        SP
        }
        \sout{
        phases, }
        \sout{we have found a step-like discontinuity 
         and features that have been proposed for the 
         Kitaev spin liquid (see 
        \ref{sec:discussion}).}
      Inset: 
        Cartoon depicting the angle-resolved torque 
          magnetometry setup.
        The arrows represent the field direction, and the 
          vector for the magnetic moment of $\alpha$-RuCl$_3$.
    }
    \label{fig:phase-diagram}
    \label{fig:axes-diagram}
\end{figure}

The Kitaev Spin Liquid ground state has limited stability 
  in the presence of competing exchange interactions, and  
  at zero field $\alpha$-RuCl$_3$ displays zigzag 
  antiferromagnetic order below $T_N \sim$ 7 K 
  (Fig.\ \ref{fig:phase-diagram}c). 
The current understanding is that a minimal description must 
  include a ferromagnetic nearest-neighbour Kitaev interaction, 
  Heisenberg interactions extending to at least the 
  third neighbours, 
  and an off-diagonal $\Gamma$ term \cite{Rau2014} that couples 
  nearest-neighbour spins (see e.g.\ \cite{Janssen2017} for a 
  discussion,  and \cite{Sears2020} for x-ray scattering results 
  determining the interaction strengths).

Above a modest field of
  $\sim7.5$ T applied in the $ab$-plane, 
  neutron diffraction shows that zigzag order is suppressed
  \cite{Sears2017}, and intense interest has focused on 
  the magnetic 
  field range $7.5 {\rm\ T} \lessapprox B\lessapprox 11 {\rm\ T}$, 
  which we refer to as the ``intermediate field phase".
This interest is enhanced by measurements 
  that are consistent with predictions  
  that a Kitaev spin liquid in
  an applied field will be a chiral 
  spin liquid, with a gap that closes for fields parallel to 
  certain high symmetry directions 
  \cite{Gordon2021,Hwang2022,Tanaka2022}.
In particular, thermal Hall measurements showed a half-integer 
  quantized thermal Hall effect \cite{Kasahara2018}, 
  which is an expected 
  signature of this phase.
Moreover, angle- and field-resolved 
  specific heat
  measurements, in the intermediate field phase 
  \cite{Tanaka2022}, were interpreted in terms of
  gapped quasiparticle excitations,  
  with a gap that closes when the applied 
  field is parallel to the Ru-Ru bonds,  
  as predicted by theory  \cite{Kitaev2006}.
(Fig.\ \ref{fig:axes-diagram}a shows a schematic of 
   the bond directions.)
Other evidence for Kitaev physics includes 
  temperature dependence of specific heat
  measurements \cite{Do2017,Sears2015,Sears2017,Widmann2019} showing 
  a predicted double-peak structure in $C(T)$, although only the 
  upper peak, near 100 K, has the predicted release of 
  $0.5\,{\rm R} \ln2$ expected for Majorana fermions.

The quantized thermal Hall effect measurements have been difficult 
  to reproduce outside a few labs \cite{Bruin2022,Kasahara2022},
  and alternative explanations have been suggested for the origin
  of the result, including phonons \cite{Lefrançois2022} 
  and topological magnons \cite{Czajka2023}.
Thus there is a clear need for 
  other measurements that could show signatures of 
  a field-induced topological spin liquid.
  
Our particular motivation for measuring the torque 
  was a prediction by some of us 
  \cite{Gordon2021}, summarized in 
  \S \ref{sec:discussion}:E, 
  that the angle-resolved torque response of a state with 
  Kitaev Ising topological order, 
  of the kind consistent with the fractional thermal Hall 
  measurements, should have a negative slope at 
  {\cgreen any in-plane C$_2$-symmetry-preserving axis,} 
  e.g.\ the B1-axis {\cgreen for a crystal with C2/m-symmetry}
  (defined in Fig.\ \ref{fig:axes-diagram}a), 
  with the slope becoming 
  more negative as $T \rightarrow 0$ K. 
This prediction is made for a generic spin model \cite{Rau2014}.

In this study we report on the magnetization of $\alpha$-RuCl$_3$ 
  through measurements of the angle-resolved torque,
  $\tau$, in applied in-plane fields of up to 9 T in the 
  temperature range of 2 - 20 K.
We have measured several single crystals and find that the 
  torque behaviour is sample dependent. 
  
{\cgreen 
There have been several recent investigations into 
   the low-temperature structure of $\alpha$-RuCl$_3$ 
   which have clarified the different $T_N$ reported 
   in the literature (from 6.5 K to 14 K)  
   \cite{Kim202403, Kim202404, Zhang2024}. 
Samples with $T_N \sim 7.5$ K have been shown to represent 
  the highest quality samples with R$\overline{3}$ symmetry 
  at low temperature,  
  minimal stacking faults and very low levels of 
  C2 contamination, while samples with $T_N<7$ K have 
  a mixture of R$\overline{3}$ and C2 regions and a high density 
  of stacking faults, while $T_N>10$ K samples have 
  ABAB stacking, as opposed to the ABCABC stacking  of 
  $T_N\sim 7$ K samples\cite{Sears2015,Banerjee2016}. }

In this paper we present results for a 
  large, \sout{high-quality,} $T_N = 7$ K crystal.
Below, and in the Appendices, 
  we justify this choice, presenting results 
  on the sample dependence of the torque, and the effects of  
  thermal cycling.

Our study complements previous torque studies of $\alpha$-RuCl$_3$,
  some of which 
  focused on the torque or the related magnetotropic 
  coefficient for out-of-plane applied fields 
  (\cite{Modic2018,Modic2021,Riedl2019}), 
  while the only previous torque study for in-plane fields that we 
  are aware of seems, in light of our results, 
  to have been 
  affected by strong residual $C_2$ strains 
  {\cgreen or disorder} 
  \cite{Leahy2017}. 

Our results are more complex than 
  predicted for a pure Kitaev interaction: 
  just outside the boundary of the zigzag phase we see a  
  sinusoidal six-fold-symmetric signal, but rather than 
  reflecting a closed gap \sout{on the honeycomb bond direction,}
  {\cgreen along the Ru-Ru bond direction,} 
  it may be indicative of short-range fluctuating zigzag order.
At low temperature, however, a 
  sharp signature with negative slope is observed at the 
  B2-axes (defined in Fig.\ \ref{fig:axes-diagram}a) 
  as soon as the intermediate phase is entered from 
  the zigzag phase.
\sout{This signature is similar to the theoretical prediction 
  \cite{Gordon2021}. }
{\cgreen
This signature's similarity to the theoretical 
  results \cite{Gordon2021} may be
  misleading, however, given its location along
  the non-C$_2$-preserving B2-axes, as well as 
  being enhanced with thermal cycling, implying extrinsic origins.}

We also report, within the zigzag phase, the existence of strong 
  sawtooth steps with six-fold symmetry and,  
  at all fields at all temperatures, a two-fold 
  torque signal that is sample and also sample-history dependent, 
  which leads us to conclude that \sout{it} {\cgreen the two-fold signal} is probably not intrinsic. 

\section{Experiment}
\label{sec:experiment}

We measured single crystals of $\alpha$-RuCl$_3$ that were 
  grown using chemical vapor transport as described 
  in previous work \cite{Lefrançois2022}. 
Two grams
  of $\alpha$-RuCl$_3$ powder (Sigma-Aldrich, Ru 45-55\%) were 
  sealed in a quartz tube under vacuum. 
The quartz tube was heated in a two zone furnace where the
  temperature gradient was kept at $70\ ^\circ$C. 
The powder was initially heated at $850\ ^\circ$C for two days, 
  followed by cooling at $4\ ^\circ$C/h down to $600\ ^\circ$C.

In all we checked the torque response of 
  14 single crystals, thirteen of which were 
  ``small", measuring $\sim 100\times50\times50\ \mu{\rm m}^3$, 
  suitable for measurement with miniature piezoresistive  
  cantilevers (Seiko Instruments PRC120). 
The most extensive set of measurements was done on 
  a ``large'' single crystal, measuring 
  $\sim 1 \times 0.6 \times 0.3$ mm$^3$, 
  with a 7 K transition (sample L7K), 
  placed on a Quantum Design Torque Chip P109A
  (Fig.\ \ref{fig:phase-diagram}b). 
For reasons presented below, and in the Appendices, 
  we believe this to be the highest quality sample, and it is 
  thus the focus of this paper.
The crystal was held in 
  position on the PPMS chip by an L-bracket 
  of copper sheet and a light layer of vacuum grease 
  (as shown in Fig.\ \ref{fig:phase-diagram}b  ), 
  with the  $c^*$-axis aligned 
  \sout{parallel to} {\cgreen within 2$^\circ$ of} 
  the axis  of rotation, \sout{and} {\cgreen which was} perpendicular 
  to the magnetic field.
This sample (L7K) has additionally been characterized 
  with x-ray diffraction to confirm its orientation. 
\sout{Measurements at low field strengths (2 T) confirmed a 
  T$_N$ of $\sim$ 7 K, in line with previous 
  literature suggesting a sample with correct ABC stacking.}

All measurements were done 
  using a Quantum Design PPMS, 
  using the Horizontal Rotator option, 
  at temperatures between 2 and 20 K, and fields from 0 to 9 T. 
In the majority of our measurements
  the crystal was rotated 
  such that the applied field swept through 
  180$^\circ$ or 360$^\circ$ in the $ab$-plane, 
  while the temperature and the magnitude of the field 
  were held constant.
{\cgreen 
These angular sweeps were done in the same direction in all
  measurements presented here. }
In all of the 360$^\circ$ studies the torque consisted 
  very precisely of two 
  repeated 180$^\circ$ patterns, as expected. 
In \sout{a few} {\cgreen most} cases we {\cgreen also} 
  swept 
temperature \sout{or field} at fixed 
  angle, to check $T_N${\cgreen
  (see Appendix \S\ref{sec:appendix-TN}). }
  \sout{or to look for quantum oscillations.}
{\cgreen 
We checked for hysteresis by 
  reversing the direction of rotation and found none within our 
  measurement resolution.}

In all of our plots, angles have been 
   translated so that 
   $0^\circ$ and $-$180$^\circ$ are at the B1 axis.
Moreover, torque is given in 
  volts (or volts per tesla) which was the signal from a 
  Wheatstone bridge that measured the twisting 
  of the sample platform.
A calibration measurement, outlined in 
  the Appendix ({\cgreen\S}\ref{sec:appendix-c}),  
  provides a conversion factor of 
  $(17 \pm 4) \times$ \sout{$10^{10}$}
  {\cgreen $10^{9}$} Nm$^{-2}$/V
  between these voltages and the torque density in Nm$^{-2}$.

{\cgreen The absence of hysteresis, 
  and calculations outlined in  
  Appendix \ref{sec:appendix-calibration}, 
  demonstrate that  the features that we observe in 
  the torque 
  arise  in the magnetic moments of our sample, 
  with minimal contamination of the signal 
  by non-linear ``torque interaction" 
  \cite{vanderkooy1968} effects.} 

\section{Results}
\label{sec:results}
Our torque signals can be decomposed 
  into oscillations with distinct periods 
  as a function of angle,  
  and in this section we discuss 
  them in sequence starting with two-fold (180$^\circ$ period) 
  signals, before turning 
  to six-fold signals which are a focus of later discussion.
\sout{We finally discuss sharp steps, or peaks, observed 
  in the intermediate field phase at low temperature near the B2-axes.}
{\cgreen We finally discuss sharp features of the ordered, 
  and then intermediate field, phases.
A special focus is put on features near high-symmetry field angles, 
  such as the B2 axis where at low temperatures in the intermediate
  field phase we see large step-like features develop.}

\begin{figure}
    \centering
    \includegraphics[width=8.4cm]{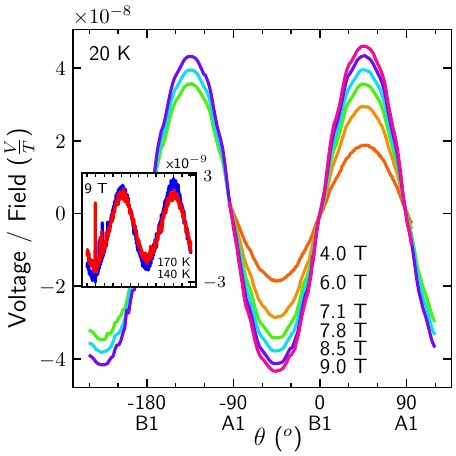}
    \caption{\justifying%
    At 20 K the torque is dominated by a 
    clear 180$^\circ$ (i.e.\ two-fold) signal that grows strongly
      with field.
    At all fields this acts as a significant background to the 
      more interesting  higher-frequency signals.
    Inset: 9 T measurements at 140 K and 170 K, showing that the
      two-fold signal persists up to \sout{and beyond} the 
      structural transition.
    {\cgreen For sample L7K the conversion from voltage to torque 
    in all plots is  2.77 ± 0.09 Nm/V, 
    while 
    the magnitude of the static moment perpendicular to the 
    applied field can be obtained from 
    $1.0\times 10^{-7}$ V/T $\leftrightarrow (0.019\pm0.001)\,\,\mu_B$ per formula unit (see Appendix \S\ref{sec:calibration} for details).
    }}
    \label{fig:2-fold}
\end{figure}

\textit{180$^\circ$ Periodic Signal:}
A clear 180$^\circ$ periodic signal appears in all of our torque 
  vs.\ angle measurements.
Fig.\ \ref{fig:2-fold} shows 
  a baseline series of measurements at 20 K at 
  fields between 4 and 9 T on sample L7K, 
  in which a 180$^\circ$ signal, that scales with
  field as \sout{$H^2$} $B^2$, dominates the torque.
\sout{Assuming that the sample is crystallographically single-domain,
  this $180^\circ$ background signal reflects 
  C$_2$ symmetry of the individual layers of 
  $\alpha$-RuCl$_3$, with a primary $a-$axis (A1) 
  that is perpendicular to the Ru-Ru honeycomb bonds, and
  $b-$axis (B1) parallel to a Ru-Ru bond 
  (see Fig.\ \ref{fig:axes-diagram}a). }
{\cgreen This 180$^\circ$ background signal reveals
  the presence of regions of our crystal with
  C2 symmetry.
(In the Discussion we argue that these regions
  coexist with R$\overline{3}$ regions.)}
The angle-resolved torque vanishes at both \sout{of these axes} 
  the B1 and A1 axes.
In all of our plots
  we have shifted the origin of the horizontal axis 
  so that B1 is at $0^\circ$ and $-180^\circ$, while 
  A1 is at $\pm90^\circ$.
As shown in the inset of Fig.\ \ref{fig:2-fold},
  the 180$^\circ$ periodic contribution 
  persists up to {\cgreen at least} \sout{, and beyond, 
  the structural transition $T_c$} $\sim 140 - 160$ K 
  \cite{Lebert2022}.
  
\begin{figure}[!htb]
    \centering
    \includegraphics[width=6.45cm]{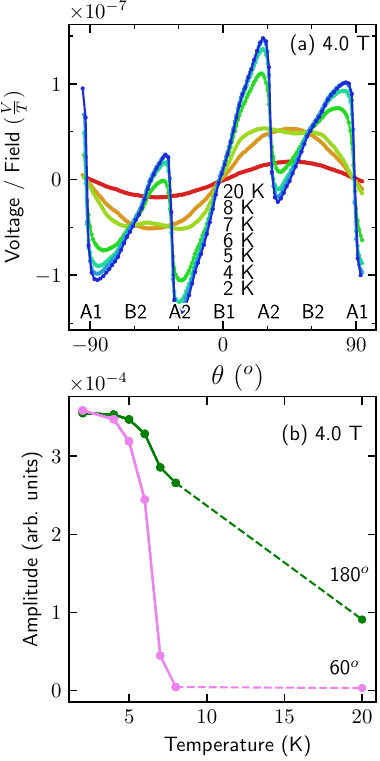}
    \caption{\justifying%
    (a) Angle-resolved torque measurements at 4 T for 
      2 K $< T < 20$ K. A nearly pure 180$^\circ$ 
      periodic signal for $T\gg T_N$ is distorted 
      starting at $\sim$ T$_N$, gaining 
      a 60$^\circ$ periodic signal which quickly transforms into a 
      sawtooth pattern. 
    (b) Fourier amplitudes of the 180$^\circ$ (green) 
      and 60$^\circ$ (pink) 
      periodic contributions to the signal in (a),  
      plotted against temperature. 
    The 60$^\circ$ periodic 
      contribution activates abruptly at $T\sim T_N$.
    }
    \label{fig:4T-composite}
\end{figure}

The 180$^\circ$ periodic signal is moderately enhanced 
  as $T\rightarrow 0$ K. 
This can be seen in Fig.\ \ref{fig:4T-composite}a, 
  which shows the angle-dependent torque at 
  4 T at various temperatures. 
Additionally, with decreasing temperature
  a pronounced six-fold (60$^\circ$) contribution appears. 
Fig.\ \ref{fig:4T-composite}b shows the temperature dependence of 
  the Fourier amplitudes corresponding to the 
  180$^\circ$ and 60$^\circ$ components. 
It is evident that these components are independent. 
From this\sout{, and further analysis in the Appendix,} we justify 
  the choice to {\cgreen individually fit and} subtract off the 
  {\cgreen temperature and  field dependent} 180$^\circ$ sinusoidal component {\cgreen for each dataset}
  in our future analyses and plots.

\begin{figure}[!htb]
    \includegraphics[scale=1.0]{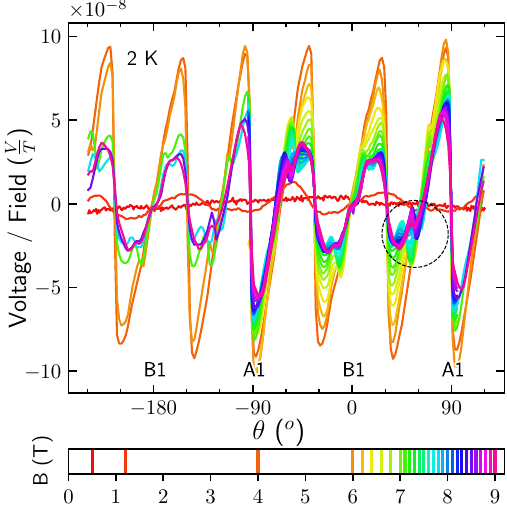}
    \caption[2K-composite-sinfit2]{\justifying%
    Angle-resolved torque of 
      L7K at 2 K. 
    A temperature-dependent 
      180$^\circ$ periodic sine wave contribution 
      has been subtracted. 
    At low field the torque response is constant, as expected.
    A 90$^\circ$ periodic contribution is present 
      at 1.2 T, and the signal further develops into a 
      60$^\circ$ periodic sawtooth signal by 4 T.
    The sawtooth discontinuities are present at each equivalent 
      $a$-axis (ie.\ $\pm30^\circ$, $\pm90^\circ$, ...). 
    The sawtooth signal distorts as field strength is increased, 
      with intermediate features developing between each 
      discontinuity, and with the overall signal 
      amplitude decreasing.
    Inflection points form along the $b$-axes (ie.\ 0$^\circ$, 
      $\pm60^\circ$, ...) with increasing field, 
      which invert across 
      the B2-axes ($\pm\ 60^\circ,\ \pm\ 120^\circ$) upon exiting 
      the zigzag phase.  
    The circled region is discussed below  
     as a possible signature of Kitaev physics 
      (see \S \ref{sec:discussion}-D). 
    }
    \label{fig:2K-composite-sinfit2}
\end{figure}

\begin{figure*}[!htb] 
    \centering
    \includegraphics[width=12.90cm]{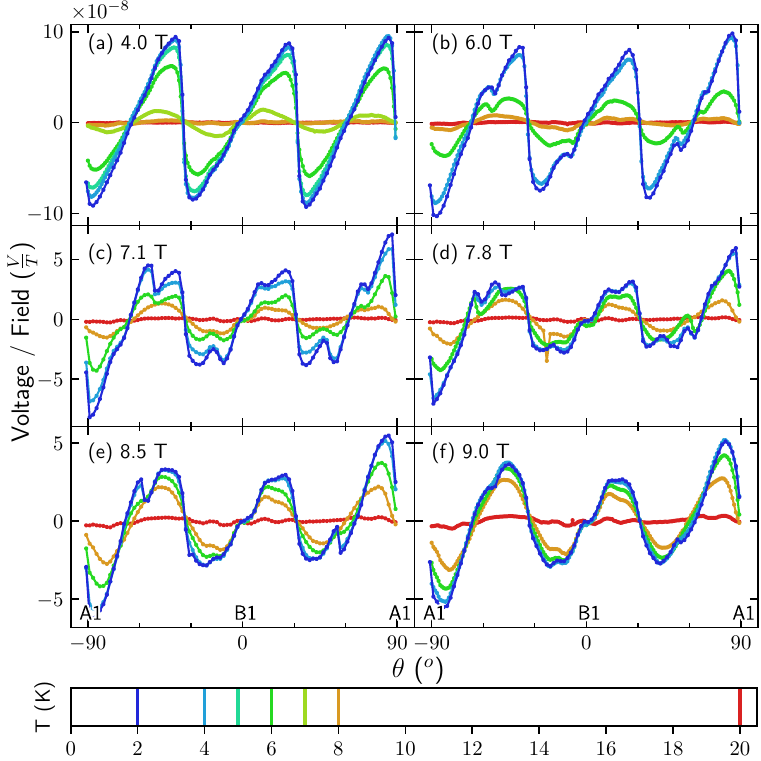}
    \caption[L7KcompositeHT]{\justifying%
    Torque vs.\ angle in sample L7K at fixed fields, 
      vs.\ temperature. 
    A 180$^\circ$ sinusoidal contribution has been subtracted.  
    (a) 4 T. At 8 K and above no features are present; 
      7 K shows a 60$^\circ$ sinusoidal contribution, without 
      harmonic distortion. 
    Below 7 K the sawtooth pattern grows rapidly. 
    (b-d) 6, 7.1 and 7.8 T. The sawtooth steps weaken 
      with increasing field, as a 
      sinusoidal pattern emerges at higher temperatures.
    A complex pattern of intermediate local extrema 
      appears. 
    The sharp step at the A1-axes ($\pm 90^\circ$)
      may be due to harmonic distortion of the two-fold 
      contribution.
    At the B1-axis (0$^\circ$) a plateau appears and grows 
      with increasing field. 
    (e-f) The step at the A1-axes, and plateau at the B1 
      axis, persist. 
    Several of the local extrema have decayed or disappeared, 
      but a sharp peak, or step-like discontinuity, emerges near B2 
      ($\sim\pm60^\circ$)  at 8.5 T.  
    }
    \label{fig:L7KcompositeHT}
\end{figure*}

\textit{60$^\circ$ Periodic Signal:} 
We turn now to the 60$^\circ$ periodic signals.
These signals are very prominent in our sample
  within the zigzag ordered 
  phase, but only for sufficiently strong fields.
  For example, as seen in Fig.\ \ref{fig:2K-composite-sinfit2}, 
  at 2 K (i.e.\ deep within the zigzag ordered phase) 
  the torque at 0.5 T, 
  after a subtraction of the two-fold contribution, is 
  nearly constant with no notable features, 
  while a field of 1.2 T induces a weak 
  90$^\circ$ periodic contribution
  {\cgreen whose origin is unclear}.  
At 4 T, however, 
  the six-fold sawtooth pattern emerges very clearly. 
In Fig.\ \ref{fig:4T-composite}b it can be seen that 
  the Fourier amplitude of the 60$^\circ$ 
  periodic component 
  is negligible for $T>T_N$, 
  but grows rapidly upon entering the zigzag phase.
Fig.\ \ref{fig:2K-composite-sinfit2} shows that 
  increasing the strength of the applied field 
  beyond $\sim 6$ T within the 
  zigzag phase distorts the signal from a clean sawtooth, with 
  the step-like features increasing in amplitude across the 
  A1-axis, while decreasing across the A2-axes.
We discuss the implication of this distortion with respect to 
  a coupling of the 180$^\circ$ and 60$^\circ$ periodic 
  contribution to the signal in \S\ref{sec:discussion}:C.
The sawtooth signal is fully suppressed at $\sim 8$ T, 
  as the intermediate field phase is entered.
This same trend can be seen at other
  temperatures T $<T_N$ in 
  Fig.\ \ref{fig:L7KcompositeHT}(a-d), where the signatures of a 
  sawtooth signal can be seen up to  7.8 T  at temperatures 
  as high as 6 K.

We argue in \S\ref{sec:discussion}:A that this sawtooth signal 
  arises from rotation of zigzag domains as the sample rotates 
  in the applied field, and that it is thus 
  an indicator of good sample quality. 
There we show that 
  the steps of the sawtooth signal are expected to 
  align with  the $a$-axes (i.e.\ A1 and A2 of Fig.\ \ref{fig:axes-diagram}a). 
We note that the orientation determined this way is 
  consistent with x-ray diffraction at room temperature
\sout{  ,but the structural transition at 
  $\sim 140 - 160$ K means 
  that additional information is needed in 
  order to orient the crystals at low temperature}. 
Locating the $a$-axes at the steps in the sawtooth,  
  together with the alignment of the zero-crossings 
  of the torque with the B1- and A1-axes at 20 K, 
  determines the orientation of our crystal at low temperature.

A novel finding of this study,
  presented in Fig.\ \ref{fig:L7KcompositeHT},
  is that a sinusoidal
  60$^\circ$ torque signal persists outside of the zigzag phase, 
  within the putative intermediate field Kitaev state.  
At 4 T a six-fold sinusoidal
  signal can be seen in a narrow temperature range below 
  7 K, within the zigzag phase, as in 
  Fig.\ \ref{fig:L7KcompositeHT}a.
At higher fields, but still with $B<B_c$, 
  the temperature range expands: the sinusoidal 
  signal now appears at higher temperatures \textit{outside} the 
  zigzag phase, as well as at lower  
  temperatures within the zigzag phase, before giving way to 
  the sawtooth signal at still lower temperatures.  
Above B$_c$,
   again \textit{outside} of the zigzag phase, 
  the 60$^\circ$ sinusoidal signal persists 
  down to the lowest temperatures measured.
  
\begin{figure*}[!htb]
    \centering
    \includegraphics[scale=1.0]{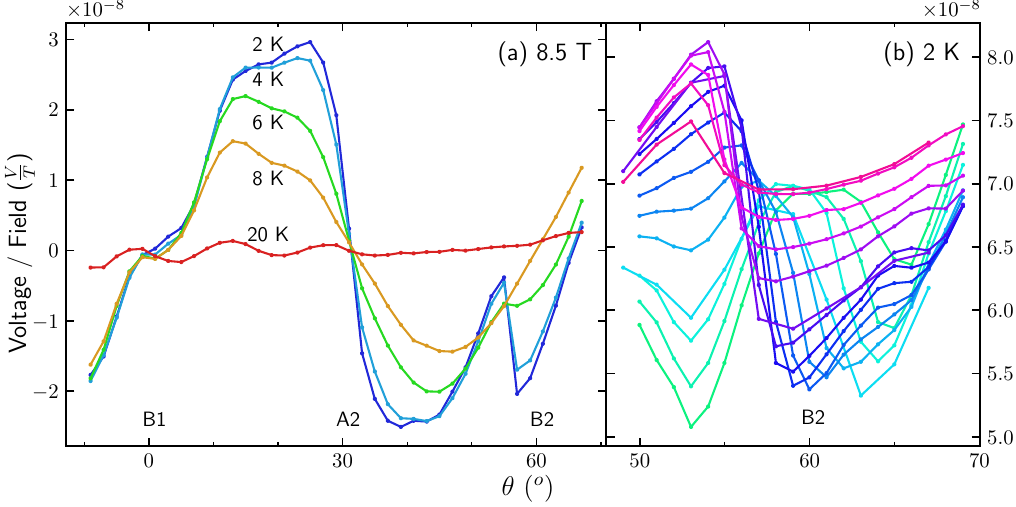}
    \caption{\justifying%
    (a) Zoomed-in window capturing the B1, A2, and B2 axes of 
      sample L7K at 8.5 T.
    A 180$^\circ$ periodic sine wave has been subtracted.
    A plateau is seen across the B1-axis, and is only 
      weakly temperature dependent. 
    A steep \sout{step-like} slope \sout{feature} develops across the A2-axis  
      (just as across the A1-axis) with decreasing temperature.
    Near B2, below a threshold temperature $\sim 4$ K, a sharp  
      {\cgreen feature with negative slope} \sout{step-like feature} develops. 
    \sout{ 
    Top inset: Angle-resolved torque about the B1 axis at 
      8.5 T and temperatures from 20 K (red) to 2 K (blue), 
      from which a further 60$^\circ$
      periodic sine wave has been subtracted.
    Bottom inset: Angle-resolved torque about the B1 axis at
      2 K and fields from 4 T (orange) to 9 T (purple), 
      from which the additional 60$^\circ$ periodic sine wave 
      has been subtracted.}
    (b) The raw (no background 
      subtraction) data of the angle-resolved torque
      near the B2-axis at 2 K,   
      for magnetic fields from 7.5 T (green) 
      to 9.0 T (purple) at evenly spaced intervals of 100 mT.
    The \sout{downward-step-like} {\cgreen steep sloped} 
      feature develops quite suddenly 
      upon entry into the intermediate field phase, and is seen 
      to move as the field changes.
      }
    \label{fig:8p5T-near-axis}
\end{figure*}

\begin{figure}[!htb]
    \centering
    \includegraphics[scale=1.0]{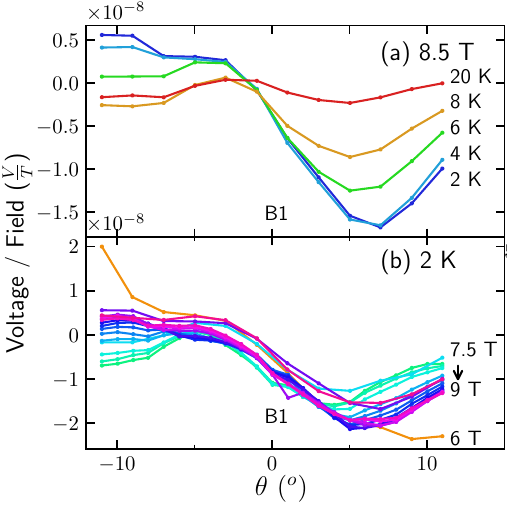}
    \caption{\justifying%
    {\cgreen
    (a)  
    Angle-resolved torque about the B1 axis at 
      8.5 T and temperatures from 20 K (red) to 2 K (blue), 
    (b)
    Angle-resolved torque about the B1 axis at
      2 K and fields at 6 T (orange), and from 7.5 T (teal) to 9 T (purple).
    In addition to a 180$^\circ$ sine wave subtraction, 
      a further 60$^\circ$ periodic sine wave 
      has been subtracted.
    }
    }
    \label{fig:8p5T-B1-axis}
\end{figure}

\textit{\sout{Features of multiple zigzag phases}
{\cgreen Other features of the zigzag phase}:}  
The transition from zigzag order to the intermediate field phase 
  is complicated by an intervening zigzag phase, 
  denoted as zigzag 2, as opposed to zigzag 1 at low field
  (see Fig.\ \ref{fig:phase-diagram}), 
  with a critical field that varies strongly with field angle 
  \cite{Balz2021,Banerjee2018}. 
In the region between about 6 and 7.7 T \cite{Balz2021} 
  where these transitions take place, we see in 
  Fig.\ \ref{fig:L7KcompositeHT}b-d 
  sharp peaks whose position is field dependent. 
\sout{A plot of the position of these extrema}
{\cgreen Plotting these features}
  vs.\ field and angle, however, reveals a pattern with little 
  similarity to the accepted phase boundaries \cite{Balz2021} 
  {\cgreen (see Fig.\ \ref{fig:field-theta-peaks} in 
  Appendix {\cgreen\S}\ref{sec:appendix-e})}. 
If the origin of these peaks is in the complex zigzag 
  phase boundaries shifting with changing in-plane field angle, 
  it is notable that both the large and small $T_N = 7$ K 
  sample host these complex features,
  though the domain rotation is seen only in sample L7K 
  (see Fig.\ \ref{fig:sample-comparison} in \sout{the}
  Appendix {\cgreen\S\ref{sec:appendix-quality}}).
{\cgreen
The placement of these peaks, between high-symmetry axes 
  instead of along them, as well as their complex dependence on 
  field strength and direction make it difficult to assess them
  further with only the measurements conducted.
However, 
  in \S\ref{sec:discussion}:B we present the possibility 
  that they arise from zigzag domain switching in C2 regions, 
  in contrast to the periodic six-fold sawtooth pattern 
  that, we will argue, 
  arises in R$\overline{3}$ regions of our sample.
In the remainder of this section 
  we focus on the peaks near 
  high-symmetry field-angles 
  in the  intermediate field phase.}

\textit{\sout{Features near symmetry axes}
{\cgreen Features of the Intermediate Field Phase}: }
In Fig.\ \ref{fig:L7KcompositeHT}d, e and f we 
  see features in the immediate 
  neighbourhood of the A1, B1, and especially, the B2 axes. 
At A1 there appears to still be a downward step, but we 
  argue below that this is due to a combination of non-linear 
  distortion of the magnetization and torque, and subtraction of 
  a pure 180$^\circ$ sinusoidal contribution 
  (see Fig.\ \ref{fig:Fint}). 
In Fig.\ \ref{fig:8p5T-near-axis}a we focus 
  on the temperature evolution of B1 and B2 features at 8.5 T.
At the zero-crossing marking the B1-axis (0$^\circ$)  
  the signal shows a plateau for temperatures below 8 K.
\sout{The insets of 
  Fig.\ \ref{fig:8p5T-near-axis}a}
  {\cgreen Figures \ref{fig:8p5T-B1-axis}a,b}
  show the signal at B1 with both a 180$^\circ$ 
  \textit{and} a 60$^\circ$ periodic sine wave subtracted, 
  showing that a negative slope in the residual signal across
  the B1-axis is enhanced with
  decreasing temperature \sout{(top inset)} {\cgreen 
  (Fig.\ \ref{fig:8p5T-B1-axis}a), but that 
  it is not field dependent for}%
  \sout{$B > 6$ T (bottom inset)}
  {\cgreen $B > 7.5$ T (Fig.\ \ref{fig:8p5T-B1-axis}b)}, 
  so we conclude that it is not a particular feature of 
  the intermediate field phase. 
  
In contrast, 
  across the B2-axes 
  a \sout{step-like discontinuity} {\cgreen steep negative slope} forms in the torque response 
  that is only seen in the intermediate field phase 
  at the lowest temperatures.
At the B2 axes at 8.5 T, for example, there is 
  an asymmetric, downward peak at $\sim 60^\circ$ 
  whose left-hand side appears to be a discontinuous 
  step, but 
  that shows up only at 4 and 2 K 
  (Figs.\ \ref{fig:L7KcompositeHT}e, \ref{fig:8p5T-near-axis}a), 
  while Figs.\ \ref{fig:8p5T-near-axis}b  and 
  \ref{fig:L7KcompositeHT}c-e show that, 
  at 2 K, the feature grows rapidly as the field is increased 
  beyond 7.5 T.
{\cgreen
This signal near the B2 axis is enhanced and shifted slightly after
  thermally cycling the sample, as seen in 
  Appendix \S\ref{sec:appendix-cycling}.}
A comparison of this feature to the theory for a Kitaev 
  spin liquid at high field is made in 
  \S\ref{sec:discussion}:D.  

{\cgreen
Finally, 
  clear oscillations, with weak temperature 
  and field dependence, appear within $\pm60^\circ$ of 
  B1 and A1 
  (see e.g.\ Fig.\ \ref{fig:L7KcompositeHT}c-f, or the 20 K line 
  in Fig.\ \ref{fig:8p5T-near-axis}a). 
These oscillations are small compared to the 
  180$^\circ$ and, 
  at low temperature, 
  the 60$^\circ$ periodic components. 
They are also weak compared to the features that develop 
   between axes in the zigzag phase, and also the 
  sharp feature which develops across the B2 axis in the 
  intermediate field phase.
These small features are distinct from the 
  temperature dependent, though field independent, 
  feature that is seen (after successive 180$^\circ$ and 
  60$^\circ$ periodic subtractions) across the 
  B1 axis.
Although we do not understand these weak oscillatory features, 
  we do not discuss them further, 
  as they are evidently inherent to 
  neither the ordered phase nor the intermediate field phase.
}

\section{Discussion}
\label{sec:discussion}
First we discuss the origin of the sawtooth pattern seen 
  within the zigzag-ordered phase 
  {\cgreen because this allows us to unambiguously determine 
  the orientation of the crystal, and the sign of the 
  torque}.  
We follow this with a brief discussion of the origin of the 
  180$^\circ$ periodic signal, 
  {\cgreen concluding that the sawtooth and two-fold signals 
  arise in physically separate parts of the sample.}
We 
  then discuss the non-sawtooth 60$^\circ$ 
  periodic sinusoidal signal which develops in proximity to the 
  antiferromagnetic region and persists into
  the intermediate field phase. 
Finally, the fine structure seen at the lowest temperatures 
  with increasing field are discussed, focusing on the features 
  that develop near the B1 and, especially, the B2 axes, 
  and their possible connection to theoretical predictions for 
  a Kitaev spin liquid.

\subsection{Interpretation of the saw-tooth pattern} 

\begin{figure}
    \includegraphics[width=8.2cm]{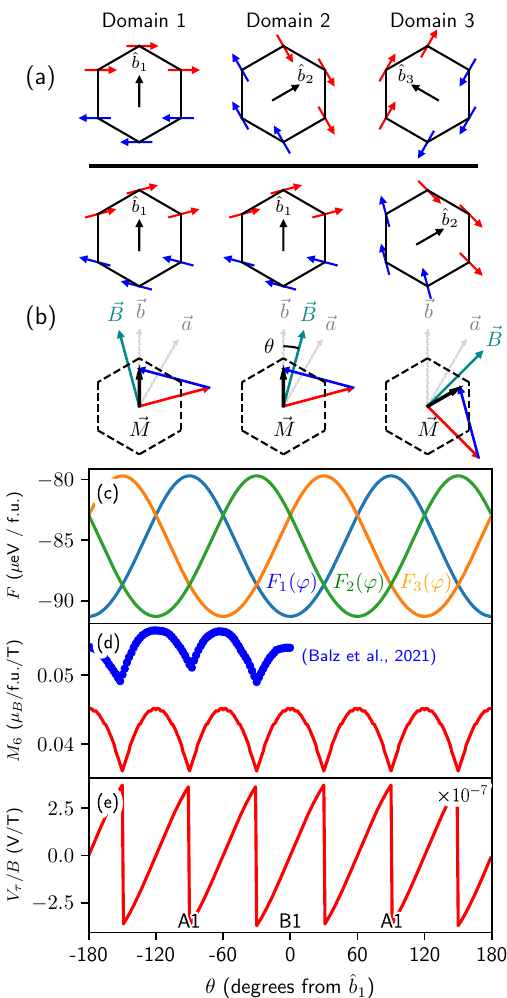}
    \caption{\justifying%
         (a) The three equivalent
           zigzag spin domains at zero field. 
           The exchange energy in our simplified model 
           depends only on the component of the magnetization 
           that is perpendicular to a Ru-Ru bond (i.e.\ to 
           a $b$-axis). 
         (b)
         In the presence of an applied field, \sout{$L_i$} 
           {\cgreen the system} switches {\cgreen domains} 
           to keep $\vec{M}_{\perp,i}$ as perpendicular to 
           the applied field as possible, minimizing the 
           exchange-plus-Zeeman energy. 
         The net magnetization, $\vec{M}$, is given by the sum of 
           the two sublattice magnetizations.
         As $\vec{B}$ passes through an $a$-axis the 
           magnetization $\vec{M}$ switches from one 
           side of $\vec{B}$ to the other, causing a step 
           in the torque.
        (c) Free energy,
        {\cgreen (d) magnetization ($M_6$),} and
        (e) torque, calculated using a toy model 
        of zigzag order as described in the text 
        (Eq.\ \ref{eq:Fzz}).
       {\cgreen 
         The filled blue circles in (d) are the digitized 
         magnetization of 
        Baltz et al.\ \cite{Balz2021}.
      For (d,e), $M_6$ and $\tau$ have 
        been converted into the units used in Ref.\ \cite{Kocsis2022}, 
        and this paper, respectively.}
          }
    \label{fig:hex}
    \label{fig:fntau}
\end{figure}

Crystal L7K shows a strong, six-fold, saw-tooth pattern in the 
  zigzag ordered phase {\cgreen (see for example the 2 K curve in 
  Fig.\ \ref{fig:4T-composite}).} 
\sout{This can be explained by the rotation of the zigzag domains to
  minimize their Zeeman energy, as is commonly observed in
  antiferromagnetic materials with C$_3$ or higher symmetry.}
\sout{We can illustrate this behaviour with a simplified free energy of 
  the form }
\ifdraft{
\begin{eqnarray}
\cancel{{\cal F}_i = KL_i^2 - \vec{M}\cdot\vec{B},}
\end{eqnarray}
} \else {} \fi
\sout{
where {\cgreen $\vec{M}$ is the net in-plane magnetization}, 
  $i=1,2,3$ labels the domain, 
  {\cgreen
  $K$ is an anti-ferromagnetic coupling constant,
  } 
  and $L_i = ( M_{\perp,i}^A - M_{\perp_i}^B)$ is the 
  zigzag order parameter, where $M_{\perp,i}^{A,B}$ is 
  the component of the magnetization on sublattice $A$/$B$ 
  that is perpendicular to Ru-Ru bond direction $b_i$. }
\sout{The spin configurations at zero field and in a rotating applied 
  field are illustrated in Fig.\ \ref{fig:hex}a,b.  At zero field 
  there are three equivalent zigzag domains 
  (Fig.\ \ref{fig:hex}a).  In an applied field,  
  the $L_i$ that minimizes the free energy is the one whose 
  sublattice magnetizations are the `most 
  perpendicular' to the applied field $\vec{B}$, 
  because this allows the spins on both sublattices 
  to have a component parallel to $\vec{B}$ while 
  still optimizing their 
  exchange energy (which favours antiparallel alignment of the 
  sublattices).}

{\cgreen If the cantilever used in a torque measurement 
  is too soft, sawtooth steps can arise 
  even when the underlying magnetization is sinusoidal, 
  a non-linear experimental artifact known as 
  ``overcritical torque interaction" \cite{vanderkooy1968}. 
We establish in Appendix \ref{sec:calibration} that 
  we are far from the overcritical regime, and thus that 
  the sawtooth steps in the torque are indeed coming from the 
  magnetization of the sample.} 
  
{\cgreen  %
The basic physics of this pattern can be understood using 
  a simplified free energy 
\begin{eqnarray}
{\cal F}_i &=& F_{zz,i} - \vec{M}\cdot\vec{B} 
\label{eq:Fzz} \\
{\rm where}\quad  
   F_{zz,i} &=& J_{zz}(\vec{M}_{A,i}\times \hat{b}_i)\cdot
              (\vec{M}_{B,i}\times \hat{b}_i) \nonumber\\
  {\rm and}\quad \qquad 
     \vec{M} &=& (\vec{M}_{A,i} + \vec{M}_{B,i}).\nonumber 
\end{eqnarray}
The subscript $i=(1,2,3)$ labels three possible orientations 
    of the zigzag domains (see Fig.\ \ref{fig:hex}(a)).
The $\hat{b}_i$ are unit vectors parallel to the 
  three Ru-Ru bond-directions, 
  and in each zigzag domain there are two sublattices, 
  A and B,
    with magnetization per formula unit (f.u.)
    $\vec{M}_{A,i}$ and $\vec{M}_{B,i}$.
The cross product with the unit vector $\hat{b}_i$ 
   thus gives the component of the
   sublattice magnetization that is perpendicular to 
   the $i^{th}$ Ru-Ru bond-direction, 
   so the $J_{zz}$ term in Eq.\ \ref{eq:Fzz} stabilizes 
   zigzag order with the moments perpendicular to $\hat{b}_i$. 

For a given direction of $\vec{B}$ within the plane, 
  we minimize the three ${\cal F}_i$'s numerically, 
  with respect to the directions of $\vec{M}_{A,i}$ and 
  $\vec{M}_{B,i}$. 
  assumed for simplicity to lie within the {$ab$-}plane.
For this calculation we fixed the magnitude of $M_A$ and $M_B$
   at $0.35\mu_B/{\rm f.u.}$,
  which is 
  representative of measured static in-plane 
  moments in the zigzag phase
  \cite{Sears2015,Johnson2015,Park2024,Cao2016,Ritter2016},
   while the zigzag exchange constant was set to 
   $J_{zz} = 1.85k_B/(0.35\mu_B)^2$, which causes the 
   zigzag order to collapse at around $B = 8$ T.

At zero field (Fig.\ \ref{fig:fntau}a) 
  the sublattice magnetizations are purely
  perpendicular to the bond, for each domain, and the
  free energies of the three domains are equal in the absence of  
  C2 anisotropy.
When a field is applied, however, 
  the sublattice magnetizations acquire a
  component that is parallel to the field 
  (upper row of Fig.\ \ref{fig:fntau}b), giving rise to a non-zero net 
  magnetization $\vec{M} = \vec{M_A} + \vec{M_B}$.
The free energies of the three domains are then no longer equal, 
  rather the free energy of each domain acquires 
  a sinusoidal dependence on the angle between $\vec{B}$ 
  and $\hat{b}_i$, as can be seen in Fig.\ \ref{fig:fntau}c.

The results are shown in Fig.\ \ref{fig:fntau}c-e. 
As the field rotates, the minimum free energy switches 
  from one domain to the next whenever the field 
  passes through an $a$-axis.
From the minimum ${\cal F}_i(\theta)$ at a given field angle 
  $\theta$ we can calculate
  the magnetization $\vec{M}$
  and the torque, 
  $\vec{M}\times\vec{B}$.
The resulting magnetization amplitude and torque are shown  
  in Fig.\ \ref{fig:fntau}d and e for $B = 3$ T. 
This field was chosen so that we could compare our 
   calculated magnetization amplitude with 
   the measured magnetization 
   in Fig.\ 3g of Balz et al.\ \cite{Balz2021} 
   (blue points in Fig.\ \ref{fig:fntau}d).
} 

\sout{The free energy of the three domains, from Eq.\ \ref{eq:Fzz}, 
  and the resulting torque, $\partial {\cal F}/\partial \theta$,  
  are shown in Fig.\ \ref{fig:fntau}c.
When the field passes through an $a$-axis, the minimum free energy 
  switches from one zigzag domain to the next. 
When this happens, 
  the magnetization vector switches from one side of $\vec{B}$ 
  to the other, causing a step in the torque.
We note that a calculation of the magnetization from this 
  simplified model closely mimics $M/B$ vs.\ angle 
  measurements at 3 and 5 T shown in Fig.\ 3g 
  of Ref.\ \cite{Balz2021}.}

{\cgreen 
In Fig.\ \ref{fig:fntau}c the domain in which
   $\hat{b}_i$ is most parallel to
   $\vec{B}$ has the lowest free energy.
This is because for this zigzag domain both 
  $\vec{M}_A$ and $\vec{M}_B$
  can have a component parallel to $\vec{B}$, lowering the
  Zeeman energy, while remaining as antiparallel to each other as
  possible, minimizing the zigzag exchange energy.
When the minimum free energy
  switches between one domain and the next, 
  the net magnetization jumps from one side of $\vec{B}$ to 
  the other (Fig.\ \ref{fig:hex}b, lower row, middle 
  and right figures), 
  producing a step in the
  torque (Fig.\ \ref{fig:fntau}e) and 
  a cusp in the magnetization (Fig.\ \ref{fig:fntau}d).
For comparison with measured data, 
  we have converted the calculated torque into the
  units used in this paper -- volts per tesla: quantitatively
  our result is about 3.5 times 
  larger than the observed sawtooth torque. 
This result is remarkably close given the simplicity of our model,
  and the assumption that our entire sample is 
  pure R$\overline{3}$ symmetry. 
  
Meanwhile 
  the magnetization is presented in units of $\mu_B$/f.u./T,
  which are the units used in Ref.\ \cite{Kocsis2022},
  for which we calculate the two-fold torque in 
  Appendix \ref{sec:misalign}.
The conversion between emu/T (used in Ref.\ \cite{Balz2021}) and
  $\mu_B$/f.u./T is approximately $9.3\times 10^{-3}$ 
  for $\alpha$-RuCl$_3$.
The similarity of our model curve and the 
  measured magnetization (``$M_6$")
  shows that the cusp in the measured magnetization
  vs.\ angle is also explained by domain rotation.
  
We note finally that Eq\ \ref{eq:Fzz} is meant only 
  as an illustration of the
  effect of zigzag domain rotation on the torque and 
  magnetization, and not as a realistic model
  for $\alpha$-RuCl$_3$.
Improved agreement with measurement can be obtained by 
  adding Heisenberg and other terms, but for illustrative 
  purposes we keep things as simple as possible.
}

Thus, the sawtooth pattern in the torque 
  is a dramatic result of 
  the switching of zigzag domains as the field rotates, 
  but this is 
  is consistent with earlier magnetization 
  measurements, which have themselves been explained by 
  more sophisticated theoretical modeling 
  (e.g.\ \cite{Janssen2017}). 
This differs, however, 
  from a recent interpretation of magneto-optical 
  spectroscopy which argued that for $B\parallel\vec{a}$ the 
  zigzag order ultimately rotates such that the spins are  
  perpendicular to that $a$-axis \cite{Wagner2022}. 
This would not produce steps in the torque. 

A sample in which the \sout{domains} 
  {\cgreen domain walls} were pinned and unable to 
  {\cgreen move} \sout{rotate} would not produce a six-fold signal in the zigzag phase
  even if all three domains were equally stable 
  at zero field; nor would a six-fold signal appear 
  in a sample with single-domain zigzag order due, 
  for example, to strong residual $C_2$ strain, 
  {\cgreen as discussed below}. 
\sout{As shown in the Appendix, 
  {\cgreen none of} our much smaller samples, having $T_N$ of 
  both 7 K (S7K) and 14 K (S14K),
  do not show six-fold sawtooth torque 
  in the zigzag ordered phase, which would be 
  consistent with either  
  of these scenarios, though which is correct is unclear.}

\subsection{Two-Fold Contribution}
All of the samples we measured 
  exhibit a strong two-fold 
  (i.e.\ a 180$^\circ$ period) signal.
The results for sample LK7 are 
  shown in \S\ref{sec:results} 
  (see Fig.\ \ref{fig:2-fold}), 
  while results for our small samples 
  {\cgreen (which tend to display 
  more pure C$_2$ symmetry in their results)}
  are shown in \sout{the} 
  Appendix \ref{sec:appendix-quality}.
Previous studies have
  assumed that 180$^\circ$ periodicity, 
  reflecting C$_2$ symmetry of the crystal, involves a shortened  
  Ru-Ru bond along a primary $b$-axis \cite{Johnson2015}, which  
  in our notation would be the B1 bond.
{\cgreen  In the zigzag phase,  
  only sample L7K shows the sawtooth torque, 
  and  a notable feature is that the sawtooth and 
  two-fold signals seem to be simply additive.} 

{\cgreen
We can add C$_2$ terms to Eq. \ref{eq:Fzz} in two ways:
1) make $J_{zz}$ larger along the B1 axis, or 
2) add an explicit C$_2$ term such as 
  $J_{C2}[(\vec{M}_A\cdot \hat{b}_1)^2 +  
     (\vec{M_B}\cdot\hat{b}_1)^2]$.
The effect on the free energy is the same. 
Fig.\ \ref{fig:Jzz_and_C2}a-c shows the free energy, magnetization 
  and torque at 3 T when $J_{zz}$ along $\hat{b}_1$ is 
  increased by a factor of 1.2, while $J_{zz}$ 
  along the other axes is unchanged. 
As can be seen, the free energy of domain 1 is lowered, 
   producing a sinusoidal two-fold torque. 
For this value of $J_{zz,1}$ 
  the free energy of the most stable domain doesn't intersect 
  that of the others, so there is no switching between 
  domains, and the 6-fold sawtooth signal is gone. 
This would appear to be a possible model for the 
  samples in which there is no sawtooth torque 
  in the zigzag phase.

\begin{figure}
\includegraphics{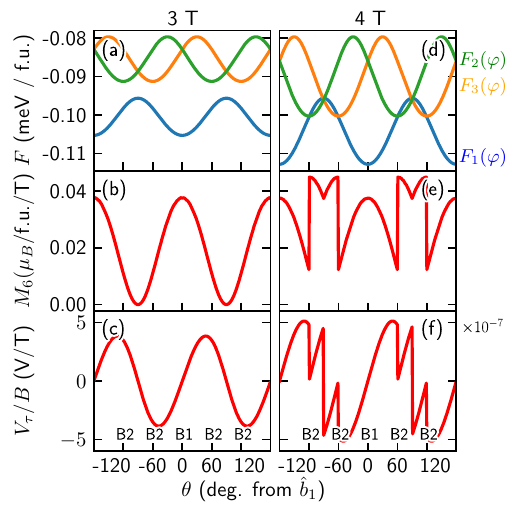}
\caption{\justifying
The effect of breaking the degeneracy between the 
  zigzag domain energies by making $J_{zz,1}$ 1.2 times 
  larger than $J_{zz,2}$ and $J_{zz,3}$. The model is 
  otherwise the same as in Fig.\ \ref{fig:hex}.
(a), (b) and (c) show the free energy of the three domains, 
  and the resultant magnetization and torque, at 
  3 T, while (d), (e) and (f) show the same quantities 
  at 4 T. 
  At 4 T the free energies now intersect at some 
  angles, causing sawtooth steps and jumps in the magnetization.
  }
\label{fig:Jzz_and_C2}
\end{figure}

In Fig.\ \ref{fig:Jzz_and_C2}d-f, 
  we have increased the field to 4 T, which increases the 
  amplitude of the sinusoidal variation in the 
  free energy to the point where the free energies 
  of domains 2 and 3 start to overlap with that 
  of domain 1. 
This causes sawtooth steps in the torque, and 
  dramatic jumps in the magnetization, but 
  they do not have six-fold 
  symmetry, 
  the steps do not occur at the $a$-axes, 
  and clearly the two- and six-fold 
  torque and magnetization are not
  simply additive.
Figs.\ \ref{fig:Jzz_and_C2}e and f look quite unlike 
  magnetization measurements \cite{Kocsis2022},
  but the steps in torque, not aligned with the A1 and A2 axes, 
  offer a possible explanation 
  of the steps that develop at intermediate 
  angles, between 6 and 9 T (Fig.\ \ref{fig:L7KcompositeHT}). 

Non-linear interference between the two-fold and six-fold 
  free energy terms, and notably moving the sawtooth steps 
  away from the a-axes, is 
  an unavoidable feature of any model where 
  these contributions 
  arise from the same set of magnetic moments, 
  and so we reach the significant conclusion that 
  the 6-fold sawtooth and the 2-fold signals in sample LK7 
  come from different regions of the sample: 
  either completely phase separated, or perhaps as 
  layered intergrowths that are intimately mixed.  }
{\cgreen This is consistent with diffraction measurements 
  \cite{Zhang2024} in $\alpha$-RuCl$_3$ that see in some 
  samples a mixture of 
  R$\overline{3}$ and C2/m regions. 
}  

{\cgreen In Appendix \ref{sec:appendix-a} it is 
   shown that the 180$^\circ$ contribution is 
   affected by thermally cycling the sample 
   from low temperature to room temperature and back, 
   while the sinusoidal 60$^\circ$ signals are not.
This suggests to us that 
   the 180$^\circ$ contribution is not  intrinsic,} 
\sout{The measurements presented in the
  Appendix suggest that
  this 180$^\circ$ periodic contribution is not intrinsic 
  to $\alpha$-RuCl$_3$ at low temperature,}
  but is instead a result 
  of residual strain 
  {\cgreen or accumulated impurity phases and stacking faults} 
  in the sample, perhaps either from 
  cooling through the $\sim 150$ K 
  structural transition too quickly, or from the way the 
  crystals are mounted, with one end stuck to the cantilever 
  with grease.
\sout{This offers a natural explanation of why, for example, 
  thermal cycling enhances the 180$^\circ$ 
  periodic contributions, without affecting the six-fold signal 
  (see Fig.\ \ref{fig:thermal-cycling-9T}, in the Appendix), 
  and why the torque in our small crystals is dominated by the 
  180$^\circ$ signal.}
Moreover, other measurements by some of us \cite{Kim202404}  
  support the notion that the {\cgreen intrinsic} 
  low-temperature  structure  is R$\overline{3}$ 
  {\cgreen (see also \cite{Zhang2024})}.  

It seems that   
  most studies on samples with predominantly R$\overline{3}$  
  symmetry 
 still show varying degrees of residual $C_2$ symmetry at 
  low temperature, and it  may be that samples have to be 
  cooled very slowly,  in strain-free conditions, 
  to show pure R$\overline{3}$ symmetry  
  (e.g.\ \cite{Balz2021,Do2017,Park2024}). 
{\cgreen 
As our measurements have already thermally cycled the sample, future
  study would be required to further explore this aspect of the low-T
  symmetry of high-quality $\alpha$-RuCl$_3$. 
}

\sout{We note that, despite the C$_2$ symmetry signal observed 
  in sample LK7, 
  evidently 
  this strain is not sufficient to prevent the rotation of 
  magnetic domains in the zigzag state.}

\subsection{Six-fold signal in the intermediate field phase}
The sinusoidal 60$^\circ$ periodic signal appears to be 
  a distinct but related phenomenon 
  from the sawtooth signal seen in the zigzag phase.
Both at the boundary to the zigzag phase at low fields, and at 
  low-temperatures in the intermediate field phase, 
  this sinusoidal signal persists where the sawtooth 
  signal is absent.
The presence of the sinusoidal six-fold signal at 4 T, 
  just below the boundary to the zigzag phase, 
  perhaps offers some insight to what the signal represents.
We know that in this region static zigzag order has set in, 
  and the existence of a six-fold signal tells us that 
  the zigzag order has a preferred domain 
  {\cgreen that depends on field direction},  
  but the sinusoidal nature of the signal tells us that
  fluctuations are still large enough 
  that all of the domains are partially populated, 
  with populations that vary smoothly as the direction of 
  the field changes.
Fig.\ \ref{fig:thermal-cycling}\sout{b}{\cgreen f}
  (in Appendix \ref{sec:appendix-a})
  shows that, at 9 T, 
  the six-fold sinusoidal signal 
  in the intermediate field phase 
  grows gradually below 20 K, before 
  saturating below $\sim 4$ K.
Given the same form of the signal just below $T_N$ at 4 T, 
  and at 9 T below 20 K, the most logical interpretation 
  is that the six-fold signal outside the zigzag phase is due to 
  fluctuating zigzag order. 

A simplified model {\cgreen of} the 
  fluctuating  
  order can be made using 
  a {\cgreen Landau} free energy with 
  anisotropic six-fold susceptiblity 
  as well as 
  C$_2$ symmetry breaking, of the following form: 
\begin{equation}
\begin{aligned}
    {\cal F}(M,\theta,\phi) 
    \simeq &
       \frac{1}{2}(\chi_\circ + \chi_2\cos(2\phi) + 
              \chi_6\cos(6\phi))^{-1}M^2  \\
    & \qquad - MB\cos(\theta-\phi), \label{eq:Fint}
\end{aligned}
\end{equation}
   where $M$ is the {\cgreen net} magnetization,
  $\theta$ is the angle of the applied field within the plane,
  $\phi$ is the angle that
  the {\cgreen net} magnetization vector $\vec{M}$ 
  makes within the plane 
  {\cgreen and the term in brackets multiplying $M^2$ is 
  $\chi^{-1}$}.  
As usual, angles are measured from the B1 axis.

For a given field $\vec{B}$, ${\cal F}$ is minimized
  with respect to $M$ and $\phi$, and 
  the torque calculated from $\tau_z = M_x B_y - M_y B_x$.

\begin{figure}
    \includegraphics[width=8.00cm]{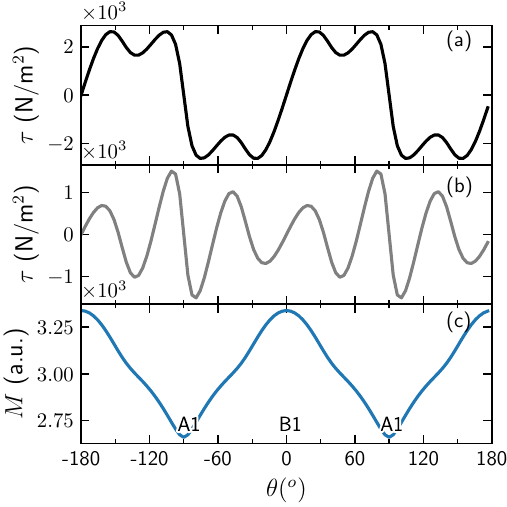}
    \caption{\justifying%
    (a) Torque density vs.\ angle for the free energy of 
      Eq.\ \ref{eq:Fint}. 
    (b) After subtracting the two-fold contribution 
      there is a downward step-like feature at $\pm90^\circ$, 
      as observed in the data at low temperature and 
      high field (see Fig.\ \ref{fig:L7KcompositeHT}e and f, 
      and \ref{fig:sample-comparison}b).
    {\cgreen
    (c) Magnetization corresponding to (a).
    } 
    } 
    \label{fig:Fint}
\end{figure}

In Fig.\ \ref{fig:Fint} we show sample results,  
  where ${\cal F}$ has been minimized numerically.
An important point is that the anomalous behaviour at the 
  A1 axis, which is apparent after a 180$^\circ$ 
  background has been subtracted (see Fig.\ \ref{fig:Fint}b) is 
  explained by non-linear coupling 
  between the $\chi_2$ and the $\chi_6$ terms, 
  which produces harmonic distortion.  
If instead of Eq.\ \ref{eq:Fint} we use a linear prefactor 
  for $M^2$ (i.e.\ of the form $(\chi_\circ^{-1} + \chi_2^{-1}\cos(2\phi) 
   + \chi_6^{-1}\cos(6\phi))$, then this anomaly at A1 is 
   not reproduced.
{\cgreen While the non-linear coupling can have many sources, 
  in this case it is probably just the non-linearity of $M$ vs 
  $B$ as the magnetization approaches saturation.}
Thus, the enhanced peaks near the A1 axis, 
  seen, for example in 
  Fig.\ \ref{fig:L7KcompositeHT}e  and f,
  are not a special feature of the six-fold physics. 
This interpretation {\cgreen requires}\sout{suggests}, incidentally, that the two-fold and at least some of the 
  {\cgreen sinusoidal} six-fold signals originate in the same regions of the crystal.

Short-range zigzag order outside of the zigzag phase is not 
  surprising.
After all, inelastic neutron scattering measurements see a strong 
  six-fold anisotropy in the magnetic scattering well above $T_N$ 
  \cite{Banerjee2017}, 
  while specific heat measurements (e.g. \cite{Widmann2019}) 
  clearly show an upturn as $T$ approaches $T_N$ from above, 
  which is a signature of fluctuating short-range order,
  {\cgreen while ESR measurements \cite{Ponomaryov17} also see six-fold variation in the ESR frequency at high field and low temperature.}  
Nevertheless, we do not have an explanation for why such 
  fluctuating order should be stronger at high field than at 
  low field, that is, why we do not see a six-fold sinusoidal 
  signal at 4 T for $T>T_N$, but we do see it at 6 T and above.
Regarding the intermediate field phase in the limit as 
  $T \rightarrow 0$ K, 
in the Introduction we mentioned 
  specific heat measurements \cite{Tanaka2022} that found 
  an apparent gap-closing six-fold anisotropy with an 
  excitation gap that closes for $\vec{B}\parallel \vec{b}$. 
This was interpreted in terms of field-induced Kitaev 
  Ising topological order, but our results suggest that 
  fluctuating zigzag order should also be taken into 
  consideration.
A possible interpretation is that, 
  because zigzag order is suppressed at 7.2 T for 
    $\vec{B} \parallel \vec{a}$, as opposed to 7.7 T for 
    $\vec{B} \parallel \vec{b}$ \cite{Balz2021}, then at 8 T, where 
    specific heat measurements
    \cite{Tanaka2022} see the strongest 
    six-fold anisotropy, 
    fluctuating zigzag order is 
    largest along the $b$-axis simply because we are closer to the 
    quantum critical point for suppression of zigzag order.

\subsection{Comparison of torque signals in the intermediate phase 
  with the theory for Kitaev Ising 
  topological order}
\label{sec:KITO}

\begin{figure}[!htb]
    \includegraphics[width=8.0cm]{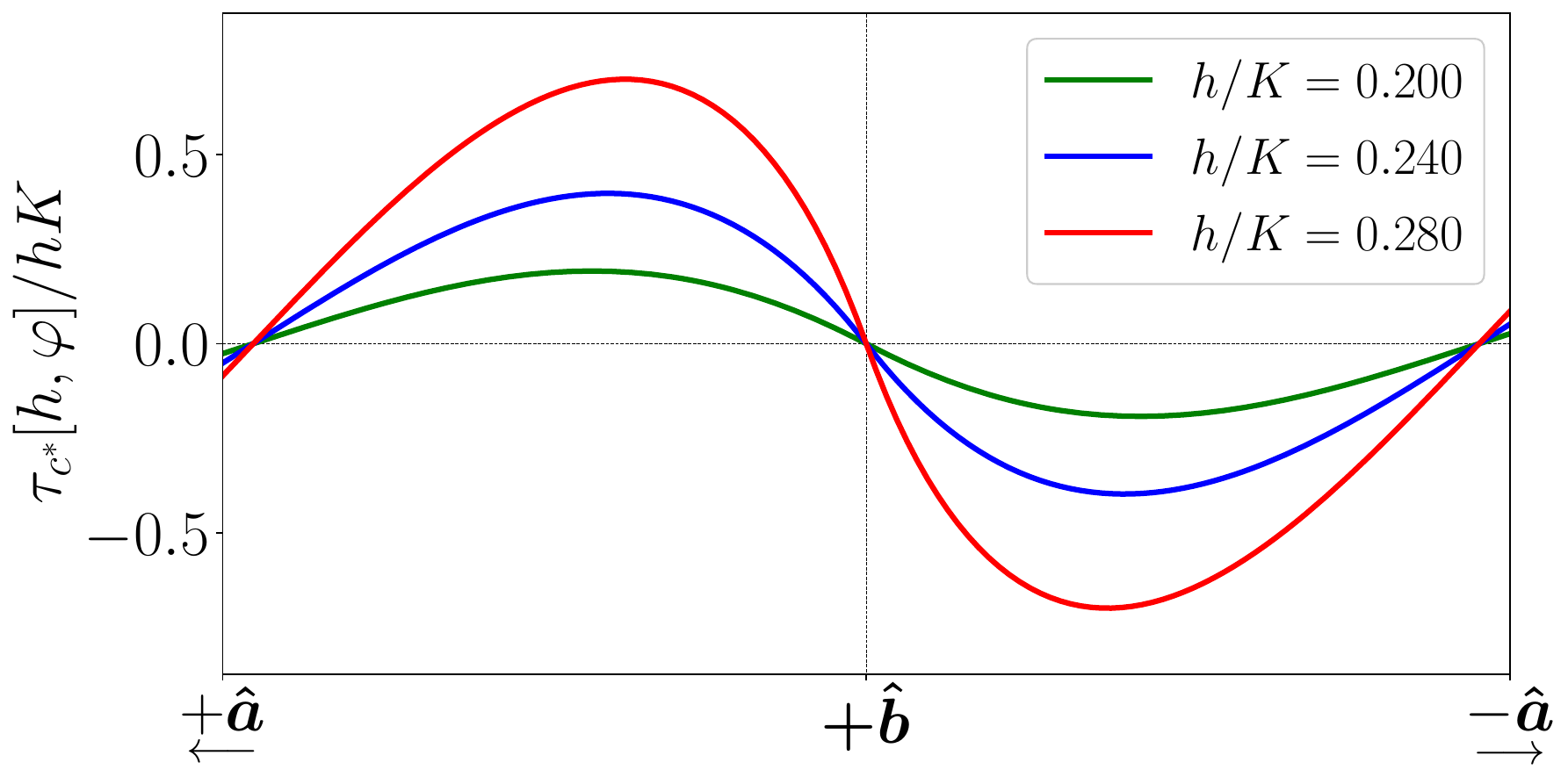}
    \caption{\justifying%
    Torque vs.\ angle calculated by 
      Gordon and Kee \cite{Gordon2021}.} 
    \label{fig:GK}
\end{figure}

\textit{Theoretical Predictions:} 
Kitaev \cite{Kitaev2006} finds that, 
  in an applied magnetic field,
  the itinerant Majorana fermions (MF's)
  of the Kitaev spin liquid become gapped,
  and they acquire a nontrivial Chern number.
The resulting phase has Ising topological order (ITO).

The ITO is prohibited when the external 
  field is parallel to a
  \sout{$b$-axis} {\cgreen C$_2$ symmetry axis} 
  since the pseudo-mirror
  symmetry is unbroken. 
As the field sweeps across such a  
  $C_2$ axis the Chern number switches from $+$1 to $-$1, 
  and
  at the axis itself the gap must close.
Gordon and Kee \cite{Gordon2021} noted that
  this topological transition leads to a sharp cusp
  singularity in the magnetotropic coefficient,
  which is independent of the details of microscopic
  spin interactions, as this special $C_2$ symmetry is a 
  property of the generic spin model. 
Writing the energy of the MF's as
  $\varepsilon_{\pm} = \pm \sqrt{(vq)^2 + m(\varphi)^2}$,
  where $m(\varphi)$ is the mass 
  and $v$ is the velocity of the MF's, 
  the free energy at $T=0$ is given by the
  integral over $q$ of the lower branch:
\begin{equation}
    \begin{aligned}
    {\cal F} &\simeq \frac{1}{2}\frac{A}{(2\pi)^2}
                    \int_{vq<\Lambda} d\vec{q}\,
                    \varepsilon_{-}(\vec{q})  \\
       &= \frac{A}{12\pi v^2}\left[ |m|^3 - (\Lambda^2 + m^2)^{3/2}
                              \right],
    \end{aligned}
\end{equation}
  where $\Lambda \gg m$ is an ultraviolet cutoff and $A$ is the
  area of the honeycomb unit cell.
Gordon and Kee \cite{Gordon2021}
  focus on the singular $|m|^3$ part of ${\cal F}$, which yields a
  cusp in the magnetotropic coefficient, 
  $\partial \tau/\partial \varphi$.
This singular part may, however, be only weakly observable 
  in the torque $\tau$, which we obtain from:
\begin{equation}
    \begin{aligned}
    \tau &= \frac{\partial {\cal F}}{\partial\varphi}%
          =  \frac{A}{2\pi v^2} \left[%
                  3m^2%
                   - \frac{3}{2}(\Lambda^2 + m^2)^{1/2}m%
                                \right]%
                                \frac{\partial m}{\partial\varphi}\\
         &\simeq -\frac{3A}{4\pi v^2}\Lambda m%
                   \frac{\partial m}{\partial\varphi}.
    \end{aligned}
\end{equation}

The mass (i.e.\ the gap) is proportional to the angle between the
  applied field, $\varphi$, and
  the $C_2$ axis, 
  $\varphi_*$: $m\propto \delta\varphi = \varphi-\varphi_*$. 
Thus, according to this calculation the torque at the C$_2$ axis
  passes through zero (as required by symmetry) but with a
  {\it negative} slope. 
The physics is relatively simple:
  the energy is lowered if the sample rotates the
  $C_2$ axis away from the field, because this
  increases the gap $m(\varphi)$, lowering the entire MF spectrum
  $\varepsilon_{-}(\vec{q})$.
This expected behaviour is shown in Fig.\ \ref{fig:GK}.

Tanaka et al.\ \cite{Tanaka2022} present strong evidence from
  specific heat for the closing of a gap at all of the
  $b$-axes in $\alpha$-RuCl$_3$, which would be consistent with 
  Ising topological order for R$\overline{3}$ symmetry.
The closing of the gap is visible below about 5 to 10 K, but
  becomes increasingly clear as $T\rightarrow 0$ K.
Thus, 
  if Ising topological order 
  is the relevant physics as opposed to, say,  proximity to the zigzag quantum critical point,
  then 
  {\cgreen in regions of our sample with R$\overline{3}$ 
  symmetry}
  \sout{if our sample has the same crystallographic symmetry as that
  used by Tanaka et al.,}
  we should expect that the torque acquires
  an increasingly negative slope at all three $b$-axes as the
  temperature is decreased below 10 K.
\sout{If,} On the other hand, {\cgreen in regions of} our 
  sample \sout{has} {\cgreen with} 
  $C_2$ symmetry 
  \sout{rather than R$\overline{3}$ symmetry,} then the 
  {\cgreen increasingly} 
  negative slope would occur only at the B1 axis. 
{\cgreen Both R$\overline{3}$ and C2 regions of the crystal 
   should contribute to a negative slope at our B1 axis, 
   that appears in the intermediate field phase and grows 
   with decreasing temperature.}
\sout{We instead see 
  a negative slope with strong temperature dependence 
  around the two B2-axis crossings, not across
  the B1-axis.} 

\textit{Experimental Results:}  
Within the zigzag-ordered phase we know that 
  the slope of the torque is positive at the $b$-axes: 
  from our model, 
  the magnetization rotates \textit{toward}
  the applied field in the zigzag phase.
On the other hand, with the nearly pure 180$^\circ$ signal 
  seen at 20 K (Fig.\ \ref{fig:2-fold}), the slope is positive 
  at the B1 axis but weakly negative at B2.

To look for a negative slope in the intermediate field 
  region at the B1 axis,
  we first subtracted  
  the strong 180$^\circ$ signal, giving the results  
  shown in 
  Figs.\ \ref{fig:L7KcompositeHT} and 
      \ref{fig:8p5T-near-axis}. 
Even after this subtraction, however, 
  we are left with the residual 
  sinusoidal 60$^\circ$ signal discussed above, which 
  has a positive slope at the B1 and B2 axes.
In light of this 
  we have also subtracted the six-fold sinusoidal 
  {\cgreen signal}
  in Fig.\ {\cgreen\ref{fig:8p5T-B1-axis}}.
  \sout{(upper inset in Fig.\ \ref{fig:8p5T-near-axis}a).} 
After this subtraction the plateaus of 
  Fig.\ \ref{fig:8p5T-near-axis}  indeed yield a negative 
  slope at the B1 axis, but the temperature 
  dependence \sout{is rather weak} 
  {\cgreen saturates below 4 K, 
  rather than becoming increasingly negative,} and the field 
  dependence is even weaker: the residual negative slope 
  persists deep into the zigzag phase 
  (Fig.\ {\cgreen\ref{fig:8p5T-B1-axis}b})
  \sout{(bottom inset of Fig.\ \ref{fig:8p5T-near-axis})}, 
  so the B1 axis torque does not appear to 
  conform to the prediction for Ising topological order. 

  The B2 axis is not trivial to locate.  We expect it to 
  be at $\pm60^\circ$, but perhaps due to interaction with the 
  $C_2$ magnetization
  \sout{, we believe from the symmetry of the signal that }
  the magnetic B2 axis \sout{is} 
  {\cgreen may be} 
  shifted to slightly lower 
  angles compared to the 
  crystallographic B2 axis.
The approximate location of the 
  isosbestic points (see Fig.\ \ref{fig:8p5T-near-axis}a and b) 
  suggests that the magnetic B2 axis, 
  in the intermediate field phase,
  is located  between 56$^\circ$ and 58$^\circ$. 

At that angle, within the zigzag phase (i.e.\ below 7.5 T) 
  the slope is 
  positive, but it rapidly becomes large and negative 
  as the field increases to 
  8.5 T, before becoming smaller up to 9 T. 
Fig.\ \ref{fig:8p5T-near-axis} shows 
  that at 8.5 T this feature corresponds to a 
  sharp downward step 
  that produces what appears 
  to be a discontinuity 
  in the torque across the B2 axis. 
Although this step is not smooth like the 
  predicted torque 
  pictured in Fig.\ \ref{fig:GK}, this may 
  be due to the limited angular resolution in our 
  measurement.  
The step has the negative sign predicted in the theory
  and it increases rapidly in 
  size with decreasing 
  temperature (see Fig.\ \ref{fig:8p5T-near-axis}a), also as predicted. 
Moreover, although this step may look similar to the 
  many peaks and steps that we see in 
  Fig.\ \ref{fig:L7KcompositeHT} between 6 and 7.8 K, 
  \sout{we believe that it is different}{\cgreen it is different 
  in that:} all of these other features fade 
  upon entry into 
  the intermediate field range, while this one is 
  strongest in the intermediate field phase, 
  just above the zigzag phase boundary. 
In this {\cgreen behaviour} it is also 
  \sout{distinct} {\cgreen different} 
  from the negative slope of 
  the (background-subtracted) signal at the B1 axis. 
Thus this downward step has many of the theoretically  
  predicted features for Ising topological order.

\sout{Finally, this step is not explainable with 
  the model we used for the steps in the zigzag phase:
  zigzag order can produce only 
  positive-slope features at the $b$-axes, 
  nor is the step explainable by 
  any other physics that we are aware of.}
  {\cgreen Fig.\ \ref{fig:Jzz_and_C2}f, however, offers another 
  explanation: this step may be due to domain switching 
  in a C2 region of the crystal with a slightly higher 
  critical field for suppression of zigzag order.
This idea is reinforced by the fact that this feature 
  grows upon thermal cycling of the sample to room 
  temperature, as does the C2 signal 
  (Appendix \ref{sec:appendix-a}).} 
\sout{Thus, while it does not exactly conform to the 
  theoretical model,  
  largely because it occurs at the B2 instead of the B1 axis, 
  the curves for fields between 8 and 9 T in 
  Fig.\ \ref{fig:8p5T-near-axis}b bear a striking resemblance 
  to the theoretical prediction illustrated in 
  Fig.\ \ref{fig:GK}, making this 
  the most promising  
  feature we have found that could be a signature of 
  the field-induced Kitaev state.}
{\cgreen
Thus, as the feature does not conform to the theoretical model, both
  by appearing along the `wrong' axis (B2 instead of B1) and 
  with the signal being enhanced after thermal cycling
  (see Appendix {\cgreen \S}\ref{sec:appendix-cycling}),
  it is likely not a signature of a field-induced Kitaev state.}
It would be worthwhile to explore it with other probes, 
  especially direct measurements of the 
  magnetotropic coefficient \cite{Modic2018,Modic2021} and 
  the magnetization, but it may be sample dependent.

\section{Conclusions}

Comparing the torque vs.\ angle between three different samples, 
across a large region of the temperature-magnetic field phase space 
has led us to following conclusions.

\begin{enumerate}

\item Two-fold anisotropy of the magnetization is both sample 
  dependent and sample-history dependent, and is probably not 
  intrinsic to the true crystallographic ground state of 
  $\alpha$-RuCl$_3$. 

\item Within the zigzag ordered phase, high quality crystals 
   show a strong six-fold sawtooth pattern in the torque, 
   that is explained by rotation of the zigzag domains to 
   minimize the magnetic interaction energy with the applied field.
   {\cgreen Modeling suggests that the sawtooth pattern arises in 
   R$\overline{3}$ regions that are spatially separated from the 
   C2 regions that generate the two-fold torque.}

\item In the high-field intermediate field phase, 
  there is a sinusoidal six-fold 
  contribution to the torque that is opposite in sign to the 
  predictions for a field-induced Kitaev state, but that may be 
  due to fluctuating zigzag order, a finding which may have 
  implications for the interpretation of zeros in the gap 
  at the $b$-axes found by specific heat 
  measurements \cite{Tanaka2022}.

\sout{ In the intermediate field phase, there are however 
  sharp step-like features in the torque at the B2 axes that 
  have many of the features predicted for 
  field-induced Kitaev Ising topological order, 
  and for which we have no alternative explanation.
  These features thus offer a promising route 
  for future investigations. }
  
{\cgreen \item A rather weak 
  negative-slope feature at the C$_2$-preserving 
  $b$-axis (B1) does not follow the temperature and field 
  dependence predicted for a Kitaev spin liquid 
  in a magnetic field.}

{\cgreen \item  We see a sharp negative-slope feature near  
  the B2 axis that does follow the predictions for a C$_2$ 
  axis of a Kitaev spin liquid in a magnetic field, but 
  B2 is not a C$_2$ axis, and moreover it and other similar 
  features at lower fields may be due to domain switching in 
  C2 regions of the crystal.
  }

\end{enumerate}

\begin{acknowledgements}
This research was supported by the 
  Natural Sciences and Engineering Research Council of Canada 
  (NSERC RGPIN-2019-06446). S. Kim and Y.J. Kim acknowledge 
  the Natural Sciences and Engineering Research
  Council of Canada (RGPIN-2019-06449, RTI-2019-00809), 
  Canada Foundation for Innovation, and Ontario 
  Research Fund. J. Gordon and H.Y. Kee acknowledge support from the Natural 
  Sciences and Engineering Research Council of Canada Discovery Grant
  No. 2022-04601. H.Y. Kee also acknowledges the Canadian Institute
  for Advanced Research, and the Canada Research Chairs Program.
\end{acknowledgements}

\appendix 

\section{Thermal Cycling Effects}
\label{sec:appendix-a}
\label{sec:appendix-cycling}

\begin{figure*}[!htb]
    \includegraphics{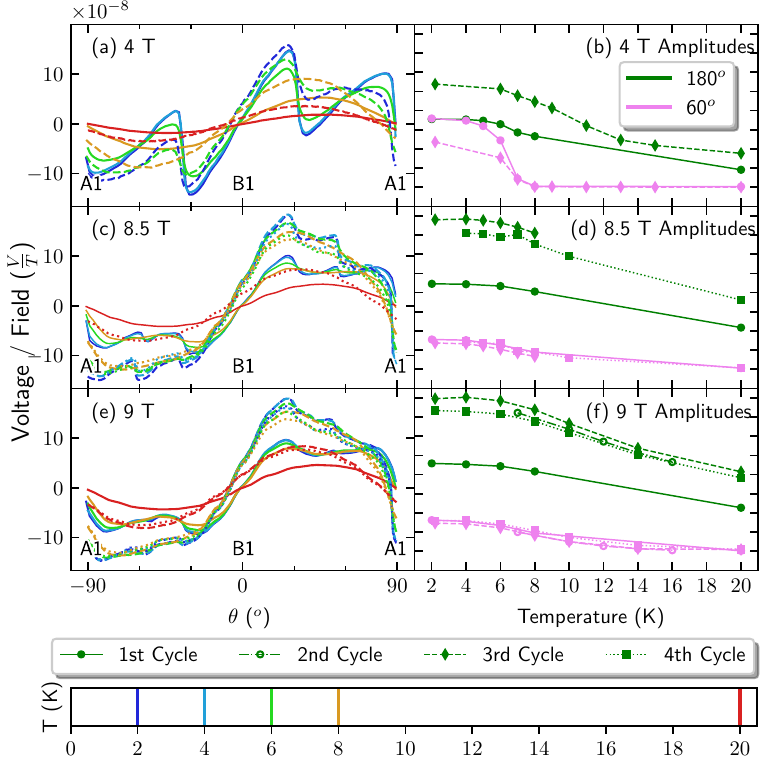}
    \caption[all-thermal-cycling] {\justifying%
    (a), (c) and (e) show the raw angle-resolved torque at 
      4, 8.5 and 9 T respectively, after successive thermal 
      cycles of the sample to room temperature and back down to 
      <20 K, measured at 2, 4, 6, 8 and 20 K. 
    (b), (d) and (f) show the 
      corresponding 
      Fourier amplitudes of the two-fold ($180^\circ$) and 
      six-fold ($60^\circ$) components of the 
      angle-resolved torque signal vs.\ temperature. 
    The two-fold component 
      of the signal is enhanced after thermally 
      cycling the sample for all field strengths, though 
      successive thermal cycles do not greatly affect the signal
      further.
    The six-fold component of the signal is not affected by
      thermally cycling the sample for field strengths outside
      of the antiferromagnetic ordered phase, though is clearly
      reduced within it (pink lines in (b)).
    }
    \label{fig:thermal-cycling}
\end{figure*}

Most of our measurements on sample L7K 
  were conducted during a single \sout{(first)} cooldown 
  from room temperature {\cgreen 
  {\cgreen (this was actually the second time the sample
  was cooled, the sample 
  having come loose during the first cooldown,  
  requiring re-mounting)}},
  but {\cgreen some} followup measurements were carried out
  after thermally cycling the sample to room temperature and 
  back to low temperature, and it was found that 
  thermally cycling sample L7K 
  {\cgreen once} enhanced the 180$^\circ$ 
  periodic contribution to the signal.
As shown in Fig.\ \ref{fig:thermal-cycling}, 
  at 7 K after thermal cycling 
  the 180$^\circ$ contribution 
  is approximately twice as large as at 8 K before cycling. 
Fig.\ \ref{fig:thermal-cycling}b shows
  the amplitude of the 
  180$^\circ$ periodic contribution, 
  extracted by Fourier analysis, as a 
  function of temperature.
It can be seen that 
  the 180$^\circ$ periodic contribution is enhanced 
  at all measured temperatures.
In contrast, a {\cgreen sinusoidal} 60$^\circ$ periodic
  contribution to the signal
  (pink line{\cgreen s} in Fig.\ \ref{fig:thermal-cycling}d,f) %
  is unaffected by thermally cycling the sample.

{\cgreen
Further thermal-cycle measurements were conducted on sample L7K to
  comprehensively observe the effects of a thermal cycling, 
  also shown in Fig.\ \ref{fig:thermal-cycling}.
  }
  
{\cgreen
Inspection of the feature at the B2 axes shows that it is 
  enhanced and shifted with thermal cycling.
This, along with the analysis of \S\ref{sec:discussion}:B 
  seen in Fig.\ \ref{fig:Jzz_and_C2}f, would suggest an origin in
  the C2 regions of the crystal.
  
We notice for these successive thermal cyclings 
  that the effect 
  on the sample seems to have saturated - further cycling does not 
  appear to noticeabley further enhance the 180$^\circ$ periodic
  contribution, and continues to not effect the 60$^\circ$ periodic
  component in the paramagnetic and intermediate field regions.
Within the zigzag phase at 4 T, however, 
  the sawtooth signal loses much of its sharpness,  
  and the 60$^\circ$ periodic contribution 
  to the signal is diminished
  (Fig.\ \ref{fig:thermal-cycling}a,b). 

A possible explanation for these observations outside of the 
  ordered phase, given the unaffected 
  60$^\circ$ component, is that 
  there exist multiple phase separated regions of the sample.
The C2 regions could be growing at the expense of the R$\overline{3}$
  regions.

} 
{\cgreen Regardless of the explanation, the lesson for subsequent 
  measurements is to measure samples with the 
  lowest possible levels of C2 contamination, 
  to minimize thermal cycling, and perhaps to cool very 
  slowly through the structural transition.}

{\cgreen
Within a given set of measurements (staying at low temperature) there
  was no sample history dependence that we observed from applying a
  field, rotating the sample in a field, or changing the temperature.}

\section{Comparisons of Sample Quality}
\label{sec:appendix-b}
\label{sec:appendix-quality}

\begin{figure}[!htb]
    \centering
    \includegraphics{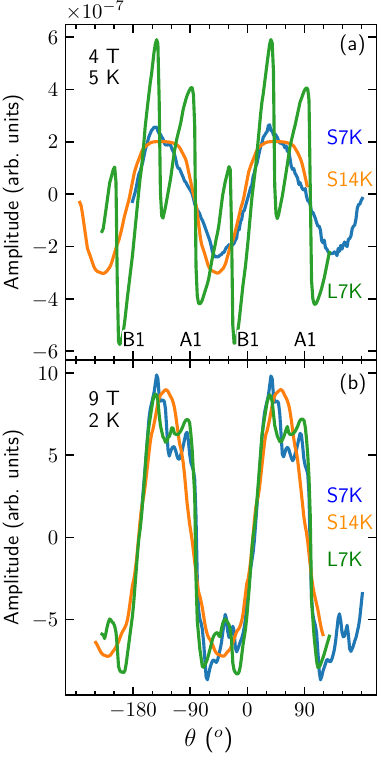}
    \caption{\justifying%
    Illustrating the differences between samples.
    (a) Torque vs.\ angle of 
      samples S7K, S14K and L7K, at 4 T and 5 K.  
    A multiplicative factor has been applied so that the 
      amplitude of the 180$^\circ$ periodic contribution is the 
      same across datasets.
    Only sample L7K shows a 60$^\circ$ periodic sawtooth, 
      superposed on the 180$^\circ$ signal. 
    (b) Angle-resolved torque response of samples 
      S7K, S14K and L7K at 9 T and 2 K. Sample S14K 
      shows only a 180$^\circ$ periodic contribution, 
      while both samples S7K and L7K 
      exhibit a complex series of higher-frequency 
      contributions superposed on a strong 
      180$^\circ$ signal.
    }
    \label{fig:sample-comparison}
\end{figure}

Over time, the understanding of sample quality has evolved 
  \cite{Cao2016,Kim202403,Kim202404,Zhang2024} to capture the effects of stacking faults, and 
  sample shape and size. 
A major distinction is between samples in which zigzag order sets 
  in at 7 K vs.\ $\sim 14$ K, which is 
  is now understood to reflect different stacking of the 
  honeycomb planes \cite{Cao2016}. 
The 7 K stacking has
  become the accepted ideal for experimental 
  measurements \cite{Kim2022}, 
  such crystals having what is believed to be the 
  intrinsic stacking (ABCABC), whereas $T_N = 14$ K 
  samples have an (ABAB) stacking.

{\cgreen
A further distinction has recently been 
  drawn \cite{Zhang2024}, 
  for 7 K samples, between those with $T_N<7$ K and those with 
  $7.2{\rm\ K} < T_N < 7.5{\rm\ K}$, the latter having a 
  sharp structural transition at 140 K upon cooling, and 
  in x-ray diffraction 
  nearly pure R$\overline{3}$ symmetry at low temperature, 
  while the former have a sluggish, broad structural transition 
  below 140 K and in x-ray diffraction 
  a mixture of R$\overline{3}$ and C2 regions 
  at low temperature. The difference is probably in 
  the density of  stacking faults,  
  as opposed to incorrectly (ABAB) stacked regions.  
}

As discussed in Appendix \ref{sec:appendix-a},
  the 180$^\circ$ periodic signal is 
  {\cgreen thus} \sout{probably} not intrinsic to the low-temperature structure 
  of $\alpha$-RuCl$_3$, but rather indicates 
  residual {\cgreen stacking faults, incorrect stacking, or} 
  strain within the sample.
We note, moreover, that small crystals seem more 
  disordered than our large crystal, regardless of $T_N$.
Both a small $T_N = 7$ K
  sample (S7K) and a small $T_N=14$ K sample
  (S14K) exhibit strong 180$^\circ$ 
  periodic behaviour (see Fig.\ \ref{fig:sample-comparison}), 
  but within the zigzag-ordered phase 
 neither sample shows the clearly defined 60$^\circ$ 
  periodic sawtooth signal of our large crystal L7K. 
A possible interpretation is that 
  finite volume effects in small 
  crystals inhibit domain rotation, although it is not clear if 
  this is due to strong $C2$ symmetry breaking, or a high 
  density of defects that pin domain walls, or both.
Similarly, Fig.\ \ref{fig:sample-comparison}(b) shows that the 
  pure 180$^\circ$ periodic signal of sample S14K persists 
  even into the `intermediate field phase', where both 
  S7K and sample
  L7K have developed additional features, including 
  the prominent {\cgreen sinusoidal} 60$^\circ$ signal
  discussed in the main body of this paper.
From this we propose that the sinusoidal 
  60$^\circ$ periodic contribution to 
  the angle-resolved torque, and by extension the presence 
  of fluctuating zigzag order persisting beyond the boundary 
  to the antiferromagnetically ordered phase, is an indicator 
  of a sample with \sout{ideal stacking} 
  {\cgreen $T_N \sim 7$ K}, although it may be that such 
  a six-fold sinusoidal signal will appear in a 14 K sample at 
  higher fields.
We note in addition that of the 13 small samples we tested, 
  12 had predominantly $T_N \sim 14$ K transitions (some crystals 
  had transitions at both 14 and 7 K), and only one had a pure 
  7 K transition. 

\sout{We note that not only sample L7K, but also 
  sample S7K, shows a sinusoidal six-fold signal in the 
  intermediate field phase, of a similar size relative to 
  the two-fold signal. 
This, together with the fact that the sinusoidal six-fold signal 
  in L7K persists up to 20 K, strongly suggests that 
  this signal is intrinsic. }

\section{Calibration}
\label{sec:appendix-c}
\label{sec:calibration}
\label{sec:appendix-calibration}
The cantilevers were measured in a low-temperature 
  Wheatstone bridge configuration, via a 4-wire measurement.
The measurements presented in this paper are given in volts,
  as measured across the Wheatstone bridge.
A series of calibration measurements, 
  with a known mass of \sout{$46 \pm 1$ mg}
  {\cgreen $4.6 \pm 0.1$ mg} of 
  solder in zero-field conditions and at 2 and 20 K, 
  was conducted 
  to derive a conversion between volts and Nm.
The mass was 
  placed $590 \pm 10$ $\mu$m off-center on the cantilever 
  to produce a twisting torque due to gravity.
{\cgreen As a function of angle}
  the calculated maximum 
  torque for this calibration mass was 
  $\tau = \vec{r} \times \vec{F}_g = $
  \sout{$(2.66 \pm 0.07) \times 10^{-7}$}
  {\cgreen $(2.66 \pm 0.06) \times 10^{-8}$} Nm.
The angle resolved torque of this calibration mass was 
  measured and the result was fit with a sin function to
  determine the amplitude.
The conversion was measured to be \sout{$27.7 \pm 0.9$ }
  {\cgreen $2.77 \pm 0.09$}
  Nm/V, which together
  with the dimensions of the sample gives a torque-density 
  conversion factor of
  $(17 \pm 4) \times$ \sout{$10^{10}$}
  {\cgreen $10^{9}$} Nm$^{-2}$/V.
{\cgreen 
At 4 T the maximum amplitude in our $\alpha$-RuCl$_3$ 
  measurements is
  approximately $1.5\times10^{-7}$ V/T, 
  which gives a magnetic moment
  of $(4.2 \pm 0.1) \times 10^{-7}$ Nm/T $= (4.5 \pm 0.1) \times 
  10^{16} \mu_B$.  
Given an estimated $16.3 \times 10^{17}$ Ru atoms, this results in
  an estimated $0.028 \pm 0.001 \,\, \mu_B$ per Ru atom for the 
  static moment, perpendicular to $\vec{B}$ and the rotation axis.
}

{\cgreen
The spring constant for the cantilever used in 
  measuring sample L7K is 
  $k=9.3\times 10^{-7} $Nm/$^\circ$.
Together with the conversion factor and approximate maximum 
  voltage measured of $8.9\times10^{-7}$ V we 
  find the maximum angular 
  deflection of the cantilever in measurements to be 
  $2.6 \pm 0.1 ^\circ$.}

{\cgreen
This maximum deflection allows us to estimate the torque
  interaction correction,
  $(\gamma/k)(\partial M/\partial \theta)B$
  for our cantilever, 
  where $\gamma$ is the deflection angle.
The condition for the torque interaction to be weak 
  is for this factor to be small compared to 
  1 \cite{vanderkooy1968}.
Even at the downward step of the sawtooth, where 
  $\partial M/\partial \theta$ is largest, 
  we find the factor to be
  quite small ($\sim 0.1$, or $10 \%$). 
  Everywhere else, in particular for purely 
  sinusoidal signals, the correction is negligible. 
This means that the sharp features we observe in the 
  angle resolved torque
  are due to the sample, and not due to non-linear
  torque interaction effects in the cantilever.
}

{\cgreen
\section{Confirming $T_N$}
\label{sec:appendix-TN}

The rapid growth of the  six-fold  torque signal below 7 K,    
   shown in  Fig.\ \ref{fig:4T-composite}b, 
   is strong evidence that the magnetic phase transition 
   took place at 7 K in sample LK7 for our first cooldown.
Here we show 
   torque at fixed angles, as a function of temperature, 
   through 12 K and 7 K for sample L7K after the 
   second cooldown (Fig.\ \ref{fig:TN}, unprimed labels)
   and after four cooldowns 
   (Fig.\ \ref{fig:TN}, primed labels). 
Although the phase transition signature is 
   not as clear, containing as they do a superposition of 
   the 6-fold and 2-fold temperature dependence,
   these plots have the advantage of 
   showing the temperature variation quasi-continuously, rather than 
   at discrete points.

\begin{figure}[!htb]
    \centering
    \includegraphics[width=8.2cm]{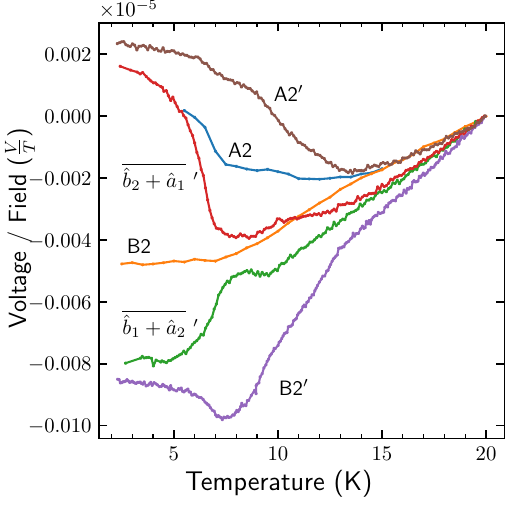}
    \caption{\justifying\cgreen%
    Measurement of the angle resolved torque of sample L7K  
      along and between high-symmetry directions,
      before and after (denoted by $^\prime$) thermal cycling,
      at a constant field strength of 2 T (before) and 
      4 T (after).
    The measurements between symmetry axes were taken half-way 
      between either the B2 and A1 axis 
      (labeled $\overline{\hat{b_2}+\hat{a_1}}'$), 
      or B1 and A2 axes, 
      (labeled $\overline{\hat{b_1}+\hat{a_2}}'$).
    The result for each dataset at 20 K is used as a subtraction
      to better contrast the different responses.
    A clear kink-like feature is present at 7 K, consistent with
      the $T_N$ seen in Fig.\ \ref{fig:4T-composite}.
    After thermal cycling a feature is seen to develop at 13 K
      along some field directions. }
    \label{fig:TN}
\end{figure}

For our first set of measurements, at 2 T the B2 curve 
  shows a change of slope in the 
  torque vs.\ temperature occurring at 7 K, 
  although there is a suggestion of a change of 
  slope at $\sim$12 K as well.
In the second, primed, set of measurements, 
  taken at 4 T and after thermal cycling,  
  the lower temperature feature is still intact, but a
  change of slope at $\sim 12$ K has become 
  more prominent at some field-angles, with a possible 
  further change at $\sim$9 K.
}

{\cgreen 
\section{Phase diagram of peaks}
\label{sec:appendix-e}
\label{sec:appendix-peaks}

\begin{figure}[!htb]
    \centering
    \includegraphics[width=8.6cm]{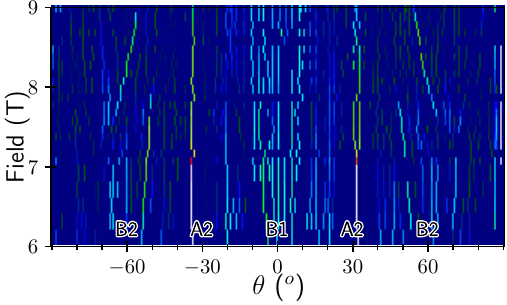}
    \caption{\justifying\cgreen%
    A composite of the peaks in the (absolute value of the)
      derivative of the angle-resolved torque 
      (ie. the magnetotropic coefficient) 
      at 2 K at cross sections of field.
    }
    \label{fig:field-theta-peaks}
\end{figure}

An analysis can be conducted of the peaks in the derivative of 
  the angle-resolved torque  
  by tracking their location in a magnetic field (T) vs.\  
  in-plane field angle ($\theta$) phase diagram
  (see Fig.\ \ref{fig:field-theta-peaks}).
These results can be compared to  the phase boundaries 
  of the zigzag 1 and zigzag 2 phases of 
  of Ref.\ \cite{Balz2021}, 
  to which our Fig.\ \ref{fig:field-theta-peaks} bears  
  very little resemblance.
This tells us that the additional steps that appear 
 between about 6 T and 8 T are not due to crossing the boundary 
 between zigzag phases 1 and 2; they are more likely due, as 
 noted above, to domain switching in regions of the crystal 
 where R$\overline{3}$ symmetry is broken.}

{\cgreen 
\section{Sample Alignment and Other Concerns}
\label{sec:appendix-f}
\label{sec:misalign}

\begin{figure}[!htb]
    \centering
    \includegraphics{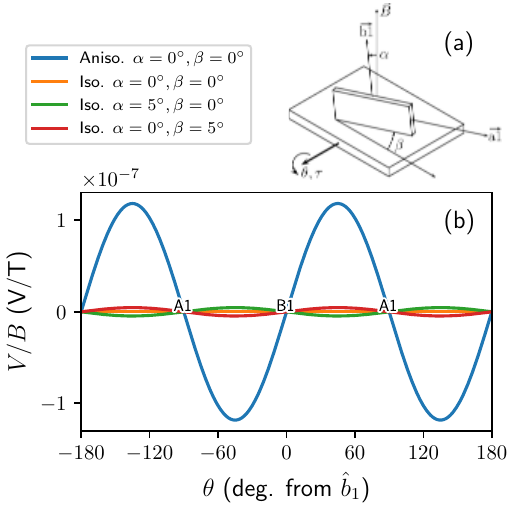}
    \caption{\justifying\cgreen%
    (a)
      Cartoon schematic of misalignment angles 
        $\alpha$ (rotation about the A1 axis), 
        and $\beta$ (rotation about the B1 axis).
    (b) 
      Calculated in-plane torque, where the anisotropic result
        utilises the measured in-plane anisotropy of 
        Kocsis \cite{Kocsis2022} with no misalignment 
        ($\alpha = \beta = 0$).
        }
    \label{fig:misalign}
\end{figure}

It was important to rule out
  the possibility that our observed two-fold anisotropy
  is due to misalignment of the sample,
  which is plausible
  given that there is approximately a factor of 10 difference 
  in $M/B$ between the in-plane and the
  $c^*$ directions in $\alpha$-RuCl$_3$ \cite{Kocsis2022}.
If the rotation axis is not precisely aligned with the $c^*$
  axis, then this large anisotropy will cause a two-fold torque
  to appear, even if the in-plane susceptibility is uniform.

We have tested this possibility numerically,
  using published $M/B$ vs $B$ data in
  the literature.
Refs. \cite{Balz2021,Kocsis2022,LampenKelley2018}
  have published $M/B$ as a function of
  field-angle, rotating the applied field 
  both within the $ab$-plane, and
  from in-plane directions to the $c^*$-axis.
All three papers find a 
  two-fold in-plane anisotropy,
  but not only the magnitude but also the sign of the anisotropy
  differs between these papers:
  in \cite{Kocsis2022} and \cite{LampenKelley2018}
  $M/B$ is largest along B1,
  but in \cite{Balz2021}, Fig.\ 3g,
    $M/B$ is largest at A1.
This striking sample dependence
  reinforces the conclusion that the two-fold in-plane
  anisotropy of $M/B$ is not intrinsic.

Our crystal L7K was oriented by eye, but the orientation was
  checked by comparing parallelism of the crystal facets
  with the sides of the rotation platform
  in a photograph of the mounted sample, and by using a microscope
  reticule and an x-y stage.
The maximum misalignment has been assessed to be less than 2 degrees.

We characterize possible misalignment
  through two angles, $\alpha$ and $\beta$ 
  (see Fig.\ \ref{fig:misalign}a).
When $\alpha=\beta=0$, the rotation axis is parallel to $c^*$,
  so that the B1 and A1 axes
  are perpendicular to the rotation axis.
For Fig.\ \ref{fig:misalign}b
  we have calculated the torque for four different cases,
  converting the calculated torque to voltage 
  using the conversion factors in Appendix C,
  and scaling by $1/B$, for comparison to our 
  measurements.
Firstly, we used the
  two-fold anisotropy from \cite{Kocsis2022} at 
  2 T and 5 T, interpolated to
  4 T for comparison with our Fig.\ \ref{fig:4T-composite}a, 
  with no misalignement (i.e.\  $\alpha = \beta = 0$; blue curve 
  in Fig.\ \ref{fig:misalign}b).
The magnitude of the $V/B$ is very similar to
  what we observed in our first cool-down 
  (see Fig.\ \ref{fig:4T-composite}a).
For the other, lower amplitude lines, we assumed
  that there is no in-plane two-fold anisotropy
  and calculated the torque:
  (i) with the sample perfectly aligned (that is, with
  B1 and A1 both perpendicular to the rotation/torque axis)
  which gives no two-fold torque (orange line);
  (ii) with $\alpha=5^\circ$, $\beta= 0^\circ$ (green curve), and
  (iii) with $\alpha=0^\circ$, $\beta = 5^\circ$ (red curve).
The torque in these cases is more than a factor of ten too
  small to explain our 
  observations, despite the misalignment angles being
  significantly larger than our estimated maximum misalignment.
Note, too, that the $\alpha$ misalignment gives the wrong sign for the
  two-fold component of torque.
We find that the $\beta$ misalignment would have to be more than 
  10 degrees for the misalignment torque to become 
  comparable in size to our measured
  two-fold torque, which is far beyond the upper 
  limit on the actual misalignment.
The misalignment torque is also too small to explain the
  change in the two-fold torque between our thermal cycling runs as
  arising from changing sample misalignment upon thermal cycling.
Thus we conclude that the two-fold signal is due to in-plane magnetic
  anisotropy that is extrinsic in origin. 
}


%

\end{document}